\documentclass{article}
\usepackage[utf8]{inputenc}
\usepackage[margin=1in]{geometry}
\usepackage[colorlinks]{hyperref}
\usepackage{graphicx}
\usepackage[labelfont=bf]{caption}
\usepackage{amsmath} 
\usepackage[style=nature, sorting=none, giveninits=true]{biblatex}
\usepackage[normalem]{ulem}

\title{\vspace{5cm}\textbf{T cell morphodynamics reveal periodic shape oscillations in 3D migration}}

\date{April 2022}

\begin{document}
\pagenumbering{gobble}

\maketitle
\begin{center}
Henry Cavanagh\textsuperscript{1}, Daryan Kempe\textsuperscript{2}, Jessica K. Mazalo\textsuperscript{2}, Maté Biro\textsuperscript{2}, Robert G. Endres\textsuperscript{1*} \\
*Corresponding author: r.endres@imperial.ac.uk
\end{center}

\noindent \textsuperscript{1}Imperial College London, Centre for Integrative Systems Biology and Bioinformatics, London, UK, SW7 2BU; \textsuperscript{2}EMBL Australia, Single Molecule Science node, School of Medical Sciences, The University of New South Wales, Sydney, Australia.

\newpage
\pagenumbering{arabic}  
\begin{abstract}
T cells use sophisticated shape dynamics (morphodynamics) to migrate towards and neutralise infected and cancerous cells. However, there is limited quantitative understanding of the migration process in 3D extracellular matrices (ECMs) and across timescales. Here, we leveraged recent advances in lattice light-sheet microscopy to quantitatively explore the 3D morphodynamics of migrating T cells at high spatiotemporal resolution. We first developed a new shape descriptor based on spherical harmonics, incorporating key polarisation information of the uropod. We found that the shape space of T cells is low-dimensional. At the behavioural level, run-and-stop migration modes emerge at $\sim$150 s, and we mapped the morphodynamic composition of each mode using multiscale wavelet analysis, finding `stereotyped' motifs. Focusing on the run mode, we found morphodynamics oscillating periodically (every $\sim$100 s) that can be broken down into a biphasic process: front-widening with retraction of the uropod, followed by a rearward surface motion and forward extension, where intercalation with the ECM in both of these steps likely facilitates forward motion. Further application of these methods may enable the comparison of T cell migration across different conditions (e.g. differentiation, activation, tissues, and drug treatments), and improve the precision of immunotherapeutic development.
\end{abstract}

\section{Introduction}

Shape changes (morphodynamics) are one of the principal mechanisms through which individual cells interact with their environment \cite{yin2014cells, bodor2020cell}. These dynamics arise from the interplay between a multitude of molecules and complex signalling pathways that often organise with emergent simplicity to carry out critical cellular functions, including division and migration. T cells, specialised cells of the adaptive immune system, are highly dependent on global morphodynamics to squeeze through gaps in the extracellular matrix (ECM), in contrast to the ECM-degrading strategies other cells use (e.g. tumour cells). Despite plasticity for adjusting the mode of migration to environmental conditions, the migration of T cells is often characterised as amoeboid: fast (up to 25 \textmu m min\textsuperscript{-1} \cite{friedl1994locomotor}), with low adhesion and polarised morphologies arising due to the segregation of different cytoskeletal networks to specific subcellular compartments \cite{weninger2014leukocyte}. In this mode of locomotion, dynamic F-actin forms pseudopods at the leading edge and an actomyosin-rich uropod at the rear generates contractile forces \cite{dupre2015t} (see Fig. \ref{fig:figure1}a for a schematic). However, this canonical migration mechanism is not fixed and T cells adapt their motility to their immediate environment.

T cells are thought to toggle between exploration and exploitation states, balancing surface receptor cues for interacting with antigen-presenting or target cells (`stop') with chemokine-driven or purely exploratory searches (`run') \cite{krummel2016t}. The specific morphodynamics and force-generating mechanisms behind these states are not well understood, in part due to their large variety and adaptability in different environments \cite{fowell2021spatio}. Proposed methods for propulsion include leading edge extension and intercalation with the ECM (using either low-adhesion integrin connections or surface texture for friction), followed by contraction of the uropod for small pore sizes \cite{soriano2011vivo, fowell2021spatio}. In addition to creating friction for moving forward, the rearward flow of actin waves from the leading edge may connect with the ECM like a paddle \cite{reversat2020cellular, abercrombie1970locomotion}. However, the extent to which these methods are used in complex 3D ECM environments, and their precise organisation, are far from well-characterised.

Accurate characterisation is important as dysregulation of T cell migration processes can be highly deleterious. T cells differentiate into different effector states. For instance, antigen-specific CD4\textsuperscript{+} `helper' T cells amplify the immune response, while CD8\textsuperscript{+} `cytotoxic' T cells seek out and neutralise infected or cancerous cells \cite{nino2020cytotoxic}. Inadequate migration leaves infected and cancerous cells free to proliferate, while over-stimulation can cause inflammation-based diseases like asthma and arthritis \cite{xia2009recent}. While there are exciting immunotherapeutic avenues manipulating the migration process, these have so far disappointed \cite{rafiq2020engineering}. With quantitative representations of T cell morphodynamics, their statistics can be interpreted with high-precision and compared across conditions for potentially improved immunotherapeutic development, and mechanistic models can be developed \cite{keren2008mechanism,tweedy2013distinct,tweedy2019screening}.

One of the main challenges for analysing morphodynamics is that cells do not have obvious landmarks (e.g. legs, eyes, wings of animals), and so the important degrees of freedom must be inferred from the data itself. Where there is important landmark-like information (e.g. polarisation that can manifest as subtle morphological features), this is typically diffuse rather than precisely-locatable, which further complicates quantification. Current methods therefore do not explicitly include this information. 2D cell morphologies are often quantified using Fourier descriptors. This method decomposes the cell outline coordinates as functions of rotation around the centroid in terms of Fourier coefficients, which then represent the morphology. This approach has revealed that amoeboid migrating cells in 2D, including epithelial keratocytes and \textit{Dictyostelium} amoebae \cite{keren2008mechanism, tweedy2013distinct}, explore only a small subspace of the shapes that might be thought possible from qualitative inspection (i.e. low-dimensionality of morphology). Furthermore, the morphodynamics within this space are composed primarily of frequently-used, or `stereotyped', motifs (i.e. low-dimensionality of morphodynamics). 

Imaging of 3D cell dynamics at sufficiently high spatio-temporal resolution has only recently become available through lattice-light sheet microscopy \cite{chen2014lattice}. Whether T cells navigating complex 3D ECM environments similarly have low-dimensional morphology and morphodynamics remains to be understood. Such questions in 3D necessitate automated analysis even more than in 2D, both because such datasets are inherently harder to visualise and interpret, and because 3D environments typically induce a richer variety of morphodynamics \cite{driscoll2015quantifying}. Spherical harmonic descriptors (SPHARM), a 3D analogue of Fourier descriptors, are a promising method for quantifying 3D cell shape and connecting with motion \cite{heryanto2021integrated}. However, the representations are typically too uninterpretable for exploring morphodynamics with high precision, and use so far is primarily limited to classification or the detection of established shape changes \cite{ducroz2012characterization, medyukhina2020dynamic}. It therefore remains an open question how best to quantify 3D cell shapes without clear landmarks and interpret high spatiotemporal dynamics.

Here, we sought to combine lattice light-sheet microscopy \cite{chen2014lattice} with quantitative image analysis to explore the 3D morphodynamics of cytotoxic T cells migrating in the absence of chemoattractant cues through 3D collagen matrices \cite{galeano2016antigen}. We first created a new compact shape descriptor, based on SPHARM, but better connected to key polarisation information than current approaches. We found that T cells explore a low-dimensional morphological space, and that run-and-stop migration emerges at long timescales. We explored the morphodynamic compositions of these two modes using multiscale wavelet analysis, previously used to explore the structure of fruit fly behaviour \cite{berman2014mapping, berman2016predictability}, uncovering a global set of largely discrete stereotyped motifs. Focusing ultimately on the run mode, due to its key role in active translocation and polarised morphologies that are well-suited for analysis with our descriptor, we found that periodically oscillating morphodynamics (every $\sim$100 s) sustain forward motion. These can be understood as a biphasic process integrating previously hypothesised propulsion mechanisms \cite{reversat2020cellular, abercrombie1970locomotion}, namely: front-widening and retraction of the uropod (rear moves forward), and rearward surface motion with forward extension (front moves forward).

\section{Results}
\subsection{T Cell Shape is Low-Dimensional}

\begin{figure}[!htb]
    \center{\includegraphics[]
    {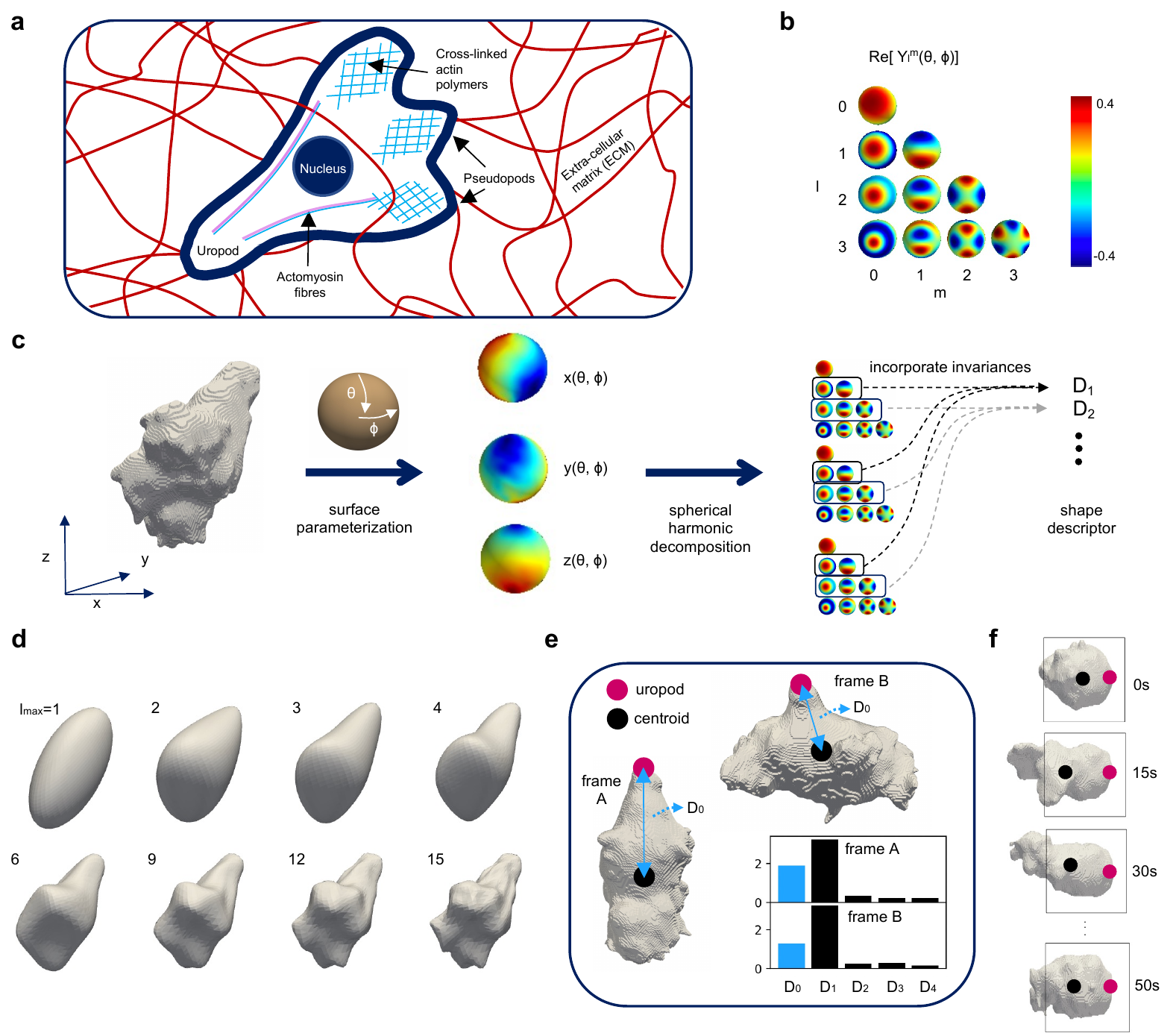}}        
    \caption{\label{fig:figure1} \textbf{T cell shape can be quantified by spherical harmonic descriptors in 3D.} \textbf{(a)} Schematic of a T cell employing an amoeboid migration strategy to navigate through the extracellular matrix (ECM) in 3D. Actin polymerisation at the front results in the formation of pseudopods, and a complex of actomyosin at the rear forms the uropod, important for stability and generating contractile forces. \textbf{(b)} Complex spherical harmonic functions, $Y_{l}^{m}(\theta, \phi)$, (real parts shown for $m\geq0$) form a basis on the surface of a sphere. \textbf{(c)} Cartesian coordinates of the cell surface, $\{x, y, z\}$, are mapped to the surface of a sphere, as parameterised by polar coordinates $\{\theta, \phi\}$. The three resulting functions $\{x(\theta, \phi), y(\theta, \phi), z(\theta, \phi)\}$ are decomposed in terms of the spherical harmonic functions and transformed to be translation, scale and rotation invariant. This yields the final shape representation, $D_{l}$, based on the harmonics at each energy level, $l$, with the exclusion of $l=0$ giving the translation invariance. \textbf{(d)} Truncation of the representation at different degrees of $l$ leads to different levels of smoothing, with $l=1$ describing the ellipsoid part of the shape. \textbf{(e)} An additional descriptor, $D_{0}$, for accounting for cell orientation, with the landmark-like smooth uropod at the rear and dynamic protrusions at the leading edge. Without this additional variable, the two cells shown have very similar descriptors. The standard deviation of $D_{0}$ across all datasets is 0.31, and the standard deviations of the remaining $D_{l}$ are all lower. \textbf{(f)} For cases where the uropod vanishes, the landmark-like rear can still be identified by its smoothness and stationarity, compared with the dynamic leading edge, as shown in the example. For simplicity, we refer to this region at the rear as the uropod for all frames.}
\end{figure}

We imaged primary mouse effector CD8\textsuperscript{+} cytotoxic T cells in 3D collagen matrices without chemical cues, with a lattice light-sheet microscope (LLSM) \cite{chen2014lattice} at spatial resolution of 0.145, 0.145, 0.4 \textmu m and temporal resolution of $\sim$2-5 s (see Methods for details on the imaging and pre-processing and Supplementary Fig. 1 for a representative snapshot and 3D trajectories). Spherical harmonics (Fig. \ref{fig:figure1}b) can be used to quantify 3D cell shapes, as shown in Fig. \ref{fig:figure1}c \cite{ducroz2012characterization, medyukhina2020dynamic, brechbuhler1995parametrization, kazhdan2003rotation}. The spherical harmonic functions, $Y_{l}^{m}(\theta, \phi)$, form a basis over the sphere, where $l$ is the function degree (related to frequency) and $m$ is the order (rotations at each degree). The full approach is detailed in Methods and summarised here. The Cartesian coordinates describing the cell surface are each mapped to a sphere, so as polar coordinates $\{\theta, \phi\}$ move over the sphere surfaces, the cell surface is traced out in object space. Analogous to a Fourier decomposition, the functions describing the cell surface can be decomposed into a set of spherical harmonic coefficients, $c_{l, i}^{m}$ with $i\in \{x, y, z\}$. The $l=0$ coefficients describe the centroid location, the $l=1$ coefficients describe the ellipsoid part of the shapes, and so on, with increasing levels of detail. Truncation of the representation at a certain $l_{max}$ therefore leads to a representation of a smoothed version of the original morphology, where higher-frequency features are filtered out (Fig. \ref{fig:figure1}d). Translation invariance is achieved by omitting the $l=0$ coefficient, scale invariance is achieved by dividing all coefficients by $V^{-\frac{1}{3}}$ where $V$ is the volume \cite{zhao2017application}, and rotational invariance is achieved by transforming to a new representation, $\{D_{l}\}_{l>0}$, with
\begin{equation}
    D_{l} = \sum_{i\in (x,y,z)}\sum^{l}_{m=0}c_{l,i}^{m}c_{l,i}^{m*},
    \label{eq:rotinv}
\end{equation}
analogous to how rotational invariance can be achieved by extracting the power spectrum from Fourier descriptors of 2D cell shapes \cite{tweedy2013distinct}. There are two key problems with the descriptor in its current form, and we made two modifications to remedy these.

\begin{figure}
    \center{\includegraphics[]
    {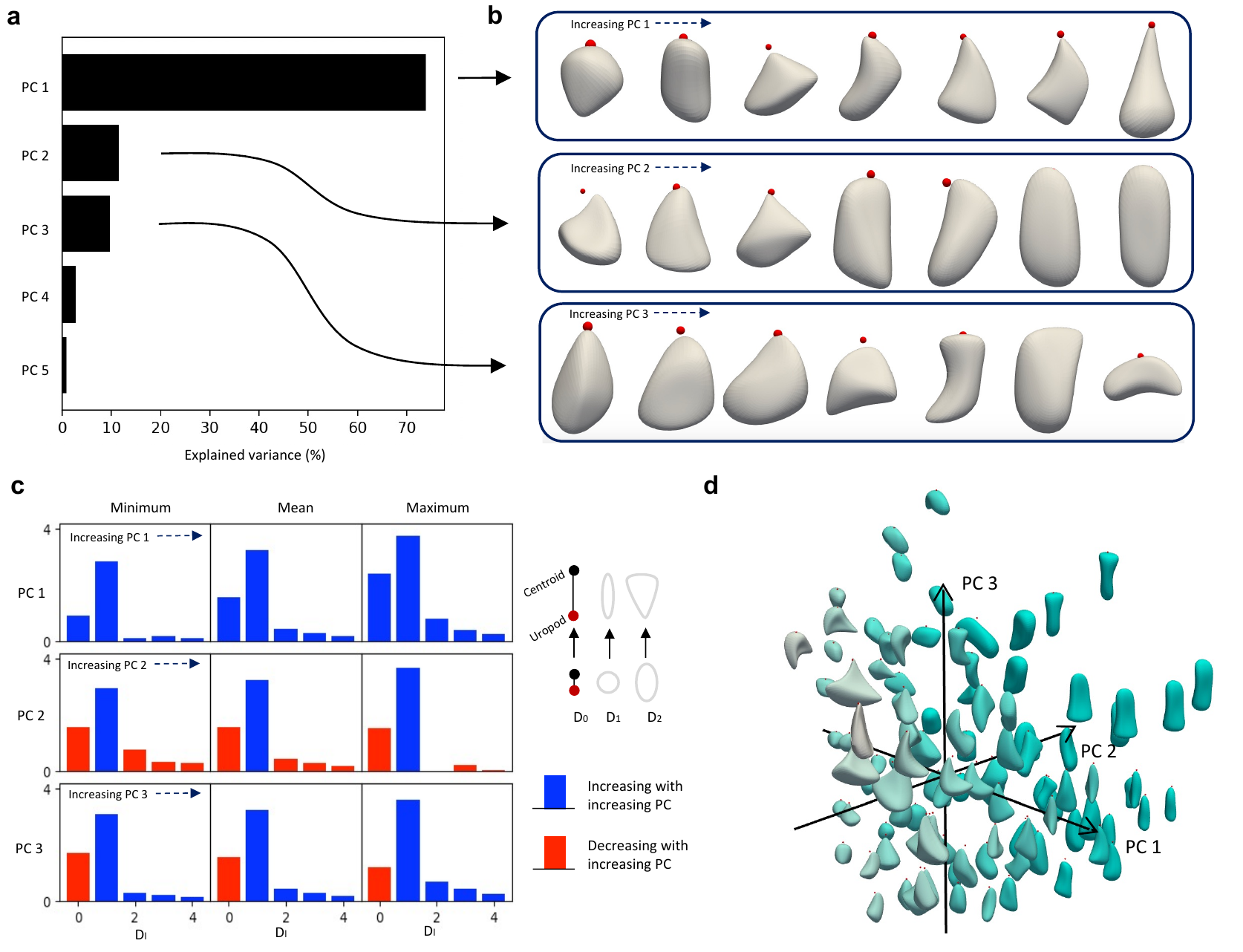}}        
    \caption{\label{fig:figure2} \textbf{T cell shape is low-dimensional as quantified with 3 principal components.} \textbf{(a)}  Principal components (PCs) 1, 2 and 3 capture 74\%, 12\% and 9.8\% (total of 96\%) of the variance in $D_{l}$, respectively.  \textbf{(b)} Shape changes associated with each PC ($l_{max}=3$ reconstructions), found by splitting the PCA space into 7 equal-length bins along each axis and plotting the T cell within each bin with the lowest value for the other PCs. An increasing PC 1 represents elongation and front-widening, a decreasing PC 2 represents contraction with front-widening, and an increasing PC3 represents elongation (forward or sideways) with the centroid moving towards the uropod. \textbf{(c)} Correspondence between the principal components (PCs) and $D_{l}$ is found by inverting the minimum, mean and maximum of each PC, with the other two PCs set to zero. Red and blue indicate decreasing and increasing descriptors, respectively, as the PCs are increased. $D_{0}$ represents the closeness between the uropod and centroid, $D_{1}$ the ellipsoidal aspects, and higher descriptors represent higher-frequency shape features. \textbf{(d)} Cell reconstructions with $l_{max}=3$ at their positions in PCA space. Darkness of colour indicates increasing PC 2.}
\end{figure}

First, the coefficients are not linearly related to the spatial extent of different features. We therefore took the square root of each element, i.e. $\{D_{l}\} \to \{D_{l}^\frac{1}{2}\}$, which yields a descriptor more representative than the power spectrum \cite{shen2009modeling}. Without this operation, almost all variance is contained in the first (ellipsoid) coefficient. Second, we added an element to the shape representation to capture key polarisation information lost in a purely global shape representation. At the cell rear is the uropod, a smooth round appendage that stabilises the cell and generates contractile forces, and at the leading edge emerge dynamic, higher-frequency protrusions. The cells in frames A and B in Fig. \ref{fig:figure1}e have very similar descriptors under a regular spherical harmonic representation, reflecting the similarity of their ellipsoid components, but this misses the polarisation conveyed in subtler features. We therefore added an extra descriptor, linearly related to the distance between the uropod and centroid, $D_{0}$ (see Methods for the full expression). The standard deviation of $D_{0}$ across all datasets is 0.31, and the standard deviations of the remaining $D_{l}$ are all lower, showing that frames A and B in Fig. \ref{fig:figure1}e are approximately two standard deviations apart along the $D_{0}$ dimension. While most cells have a well-defined uropod that can be readily identified (e.g. frames A and B in Fig. \ref{fig:figure1}e), some can exhibit more spherical shapes, as shown in Fig. \ref{fig:figure1}f. However, even for these cells there is still an identifiable smooth rear opposite a dynamic leading edge, and temporal information can reveal where the uropod transiently forms. For simplicity, we refer to this region at the rear as the uropod for all frames. The ultimate representation of T cell shape is therefore $\{D_{l}\}_{l=0}^{l_{max}=15}$ with $D_{0}$ as described above and $D_{l}$ for $l>0$ the square root of the expression in Eq. \ref{eq:rotinv}.

We used principal component analysis (PCA) to identify a set of uncorrelated linear features, or principal components (PCs), from the initial high-dimensional shape representation, $\{D_{l}\}$. Despite the lack of obvious constraints from manual inspection, Fig. \ref{fig:figure2}a shows only three PCs are required to capture $\sim$96\% of the variance in the data (74\%, 12\% and 9.8\% for PCs 1, 2 and 3, respectively). The rotational invariance means that the PCA coordinates are not invertible to unique shapes. To better isolate what features each PC describes, we therefore split the PCA space into 7 equal-length bins along each axis and plotted the T cell within each bin with the lowest value for the other PCs, shown in Fig. \ref{fig:figure2}b for $l_{max}=3$ reconstructions and Supplementary Fig. 2a for full cells (and Supplementary Fig. 2b shows the PC values of these plotted cells).  Fig. \ref{fig:figure2}c shows what $D_{l}$ transitions these PCs correspond to, with the minimum, mean and maximum inverted for each PC (with the other PCs set to zero), and Supplementary Fig. 2c shows the vector composition of each PC. An increasing PC 1 represents elongation and front-widening, a decreasing PC 2 represents contraction with front-widening, and an increasing PC3 represents elongation (forward or sideways) with the centroid moving towards the uropod. Fig. \ref{fig:figure2}d shows a sample of cells (with $l_{max}=3$) at their locations in the PC space. Supplementary Fig. 2d shows that along the main axis of variation (PC 1), dimensionality is relatively constant, and Supplementary Fig. 2e shows only modest differences in the spherical harmonic spectra of the low and high PC 1 populations.

Uncertainty in the uropod label, a diffuse region rather than a precisely-locatable point, can be quantified and propagated to downstream variables of interest (Supplementary Fig. 3 and Methods). Uropod uncertainty was found using the curvature around the labelled point (Supplementary Fig. 3a-b), and then PC uncertainties were calculated by re-computing $D_{0}$ using each point on the cell rear within this uncertainty (Supplementary Fig. 3c). The mean percent uncertainty in $D_{0}$ is 1.5\%, which is lower than the uropod uncertainty since cell rears are typically perpendicular to the axis defined by the centroid and uropod. The percentage uncertainties of the PCs (relative to their standard deviations) are 4.4\%, 0.30\% and 9.2\% for PCs 1-3, respectively.

\subsection{Run-and-stop migration emerges over long timescale}

\begin{figure}
    \center{\includegraphics[width = 0.7\linewidth]
    {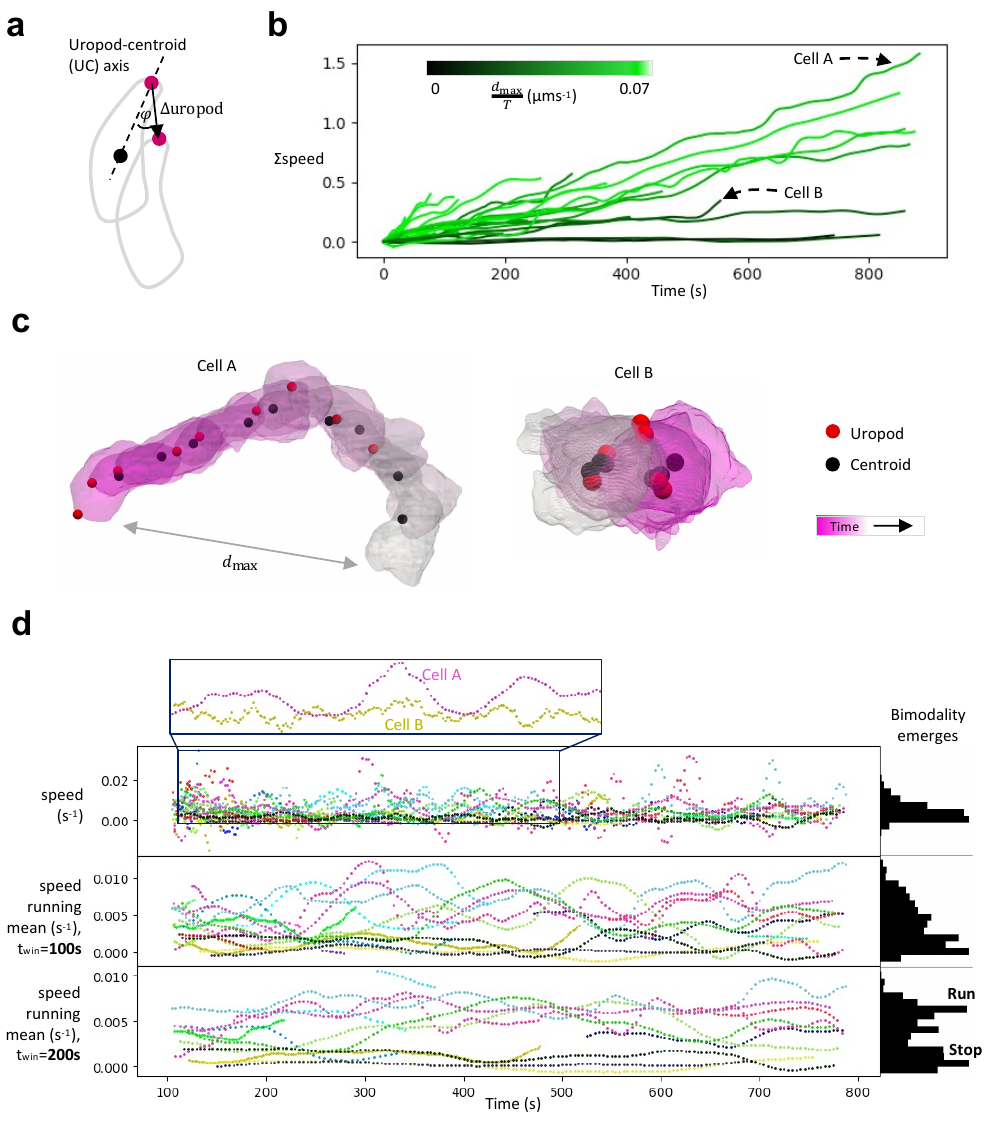}}        
    \caption{\label{fig:figure3} \textbf{Run-and-stop migration emerges over long timescales.} \textbf{(a)} Speed is defined as the uropod speed along the uropod-centroid (UC) axis, $\| (\Delta\textnormal{uropod}/\Delta t) \|\cos{\varphi}$, with smoothed uropod and centroid locations, and a further operation for invariance to cell scale (see Methods). \textbf{(b)} Cumulative speed plots show some cells have repeated phases of high speed (e.g. Cell A) while others have much lower speeds (e.g. Cell B). Lines are coloured by the maximum distance traveled divided by total dataset duration. \textbf{(c)} This is despite significant uropod motion in some cases. Meshes are shown every $\sim$104 s and 103 s for Cells A and B, respectively. \textbf{(d)} Histograms of speed with different running mean windows ($t_{win}$). At small timescales, differences in speed between cells can be indistinguishable because most exhibit phases of low speed, highlighted for Cells A and B. However, bimodality into two modes (run-and-stop) emerges at around 150 s.}
\end{figure}

To connect morphodynamics with migration strategies, variables describing cell motion are required. There are two landmark-like features of the cell that move through the ECM, the uropod and the centroid, and we calculated velocity vectors for both, invariant to cell scale (i.e. units of s\textsuperscript{-1}; see Methods). To ensure uropod velocities have adequate signal-to-noise ratio (SNR), where noise arises from uropod labelling uncertainty, we found for each dataset the mean time taken for the uropod to move a significant distance, $\tau_{sig}$, and then computed velocities using running means over position with a time window of  $\tau_{sig}$ (see Methods and Supplementary Fig. 3d for details). We then calculated speeds, which are 1D and rotationally-invariant, unlike velocities. The uropod and centroid speeds alone cannot separate distinct behaviours at small timescales, like translation and rotation (Supplementary Fig. 4a), and so we searched for a biologically-meaningful reference frame. We found that long-timescale migration is typically along the axis defined by the uropod and centroid (the UC axis), rather than the ellipsoid major axis (Supplementary Fig. 4b). The speeds of the uropod and centroid along this axis then better differentiate distinct motifs (Supplementary Fig. 4c) and Supplementary Fig. 4d shows these describe largely irreversible motion. The former has lower variance and fewer reversals (Supplementary Fig. 4d), and Fig. \ref{fig:figure1}f and Supplementary video 9 show dynamics that we observed in some datasets, where the cell appears to test routes with multiple extensions and retractions but a relatively static uropod, before committing with the uropod. We therefore selected uropod speed along the UC axis as the variable for cell motion (Fig. \ref{fig:figure3}a), and henceforth refer to it simply as speed.

Two migration modes separate out at long timescales, as shown in plots of cumulative speed (Fig. \ref{fig:figure3}b): repeated phases of high speed, making significant progress forward, e.g. Cell A; and lower speeds, yielding little progress, despite significant uropod motion in some cases, e.g. Cell B (Fig. \ref{fig:figure3}c). Fig. \ref{fig:figure3}d shows that, while at small times the dynamics can be indistinguishable (both modes have phases of near-zero speed), run-and-stop bimodality emerges at approximately 150 s. This bimodality is consistent with conclusions from lower-resolution experiments, where long-timescale trajectories of single cells have been modelled with Lévy-type random walks (characteristic of switching between stop and run modes) \cite{harris2012generalized}. Interestingly, another study suggested more complex statistics, with cells divided into sub-populations described by distinct random walk models \cite{banigan2015heterogeneous}. PCs 1 and 2 have a stronger correlation with run-and-stop mode than speed, indicating that shape is specialised more for migration mode than instantaneous speed, with cells in the run mode longer and thinner than those in the stop mode (Supplementary Fig. 5a). We next explored the morphodynamics behind these migration modes.

\subsection{Stereotyped morphodynamics underlie migration modes}

We analysed longer duration datasets for each of the run and stop modes to investigate how they differ (Supplementary videos 1-4 and 5-8 for the run and stop modes, respectively). We first computed the autocorrelation functions of the shape (PCs 1-3) and speed dynamics (using high SNR timeseries; see Methods for details). The autocorrelation function (ACF) is the correlation of a timeseries with a lagged version of itself, as a function of the lag, which can reveal the presence of latent variables preserving information across time.  We found an autocorrelation decay time, $\tau\textsubscript{ACF}$, by fitting an exponential decay model to the peaks of the oscillating ACFs (Supplementary Fig. 5b), and these decay times are indicative of the timescales over which processes are likely guided more by internal cytoskeletal machinery than stochastic external cues. For the stop mode, PC 3 is more autocorrelated than the other variables (mean $\tau\textsubscript{ACF} \sim$250 s compared with $\sim$150 s of the other variables; Supplementary Fig. 5b). PC 3 dynamics are suggestive of sensing: forward extension with a tentative rearward centroid, and reaching sideways. See Supplementary videos 5-9, with the three included in the PC 3 ACF analysis coloured by PC 3. For the run mode, the main differences are a decrease in the PC 3 autocorrelation (to $\sim$150 s) and an increase in the speed and PC 2 (contraction with front-widening) autocorrelations (to $\sim$225 s). The power spectra in Supplementary Fig. 5c show the run mode has larger oscillations in speed and PC 2, particularly for 0.005-0.01 Hz. The run mode is therefore associated with faster oscillations in speed and PC 2 that typically remain autocorrelated for longer than those of the stop mode. These ACFs give a global perspective on morphodynamics, and the presence of long timescales suggest that the morphodynamics are, as with the morphologies, low-dimensional. We therefore next zoomed in on the PC timeseries to interpret the organisation of local morphodynamics, or `behaviours', that underlies this low-dimensionality.

\begin{figure}
    \center{\includegraphics[]
    {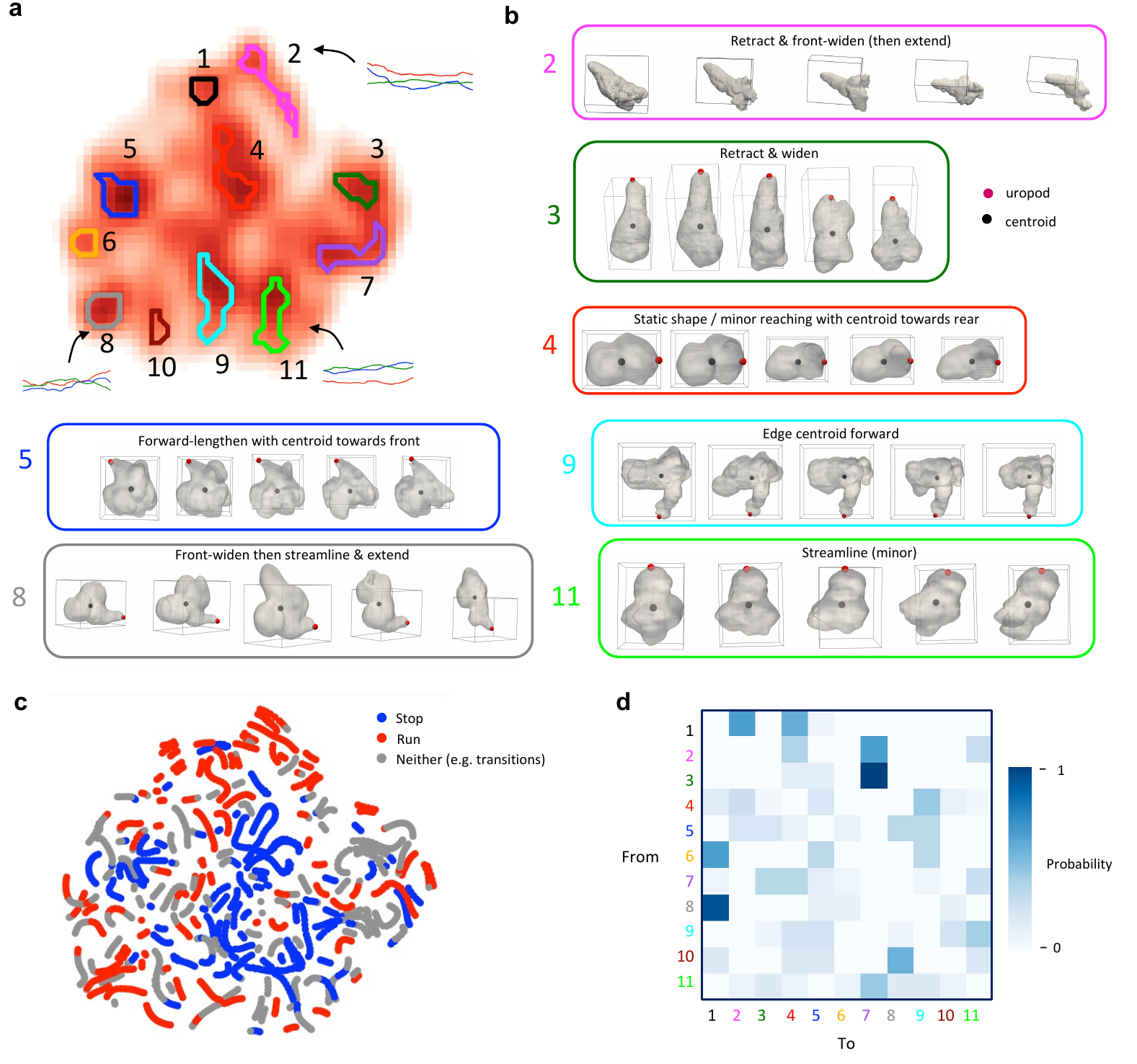}} \caption{\label{fig:figure4}\textbf{Morphodynamics are organised in stereotyped motifs.} \textbf{(a)} 11 stereotyped motifs are local peaks in a probability density function (PDF) over the spectrogram embeddings in morphodynamic space, where different locations represent different local morphodynamics. These form more of a discrete set than continuum, and some examples of the local PC series are shown (red, blue and green lines for PCs 1-3, respectively). \textbf{(b)} Example stereotyped motifs, with frames evenly-spaced across 150 s. \textbf{(c)} Utilisation of motifs in the stop and run modes. Red and blue indicate the speed running mean is above 0.005 s$^{-1}$ (run mode) and below 0.0025 s$^{-1}$ (stop mode), respectively (selected from the bimodal distribution in Fig. \ref{fig:figure3}d), and grey indicates it is in between these values (e.g. transitions). \textbf{(d)} Transition probability matrices reveal how cells move around the morphodynamic space, counting transitions only once the cell moves to a different motif.}
\end{figure}

The continuous wavelet transform is a method for finding local morphodynamics (behaviours) from a timeseries of morphologies, and has been used to map stereotyped behaviour in fruit flies \cite{berman2014mapping, berman2016predictability}. See Methods for the full pipeline and Supplementary Fig. 6a for a schematic. Wavelets are used to transform the timeseries into a spectrogram with multiscale dynamic information. Dimensionality reduction with t-SNE \cite{van2008visualizing} can then be performed to map the spectrogram to an interpretable 2D morphodynamic space, where different locations represent different local morphodynamics (Fig. \ref{fig:figure4}a), and Supplementary Fig. 6b shows the dimensionality reduction is robust across different hyperparameters. Stereotyped motifs are those that are frequently performed, and so correspond to peaks in the probability density function (PDF) of spectrogram embeddings in this space. We used wavelets with a maximum width of influence of 150 s, the approximate timescale of organisation found from the autocorrelation analysis.

We found that behaviours are organised into more of a discrete set rather than a continuum (Fig. \ref{fig:figure4}a): `islands' between which cells jump, and we could therefore categorise and interpret these individually. Fig. \ref{fig:figure4}b shows key examples, with frames evenly-spaced over a 150 s interval (with the remainder and further examples in Supplementary Fig. 7), and Supplementary Fig. 8 shows the PC dynamics of three examples from each motif. Fig. \ref{fig:figure4}c shows how these are utilised differently in the run and stop modes. Red and blue indicate the speed running mean is above 0.005 s$^{-1}$ (run mode) and below 0.0025 s$^{-1}$ (stop mode), respectively (selected from the bimodal distribution in Fig. \ref{fig:figure3}d), and grey indicates it is in between these values (e.g. transitions). In the stop mode, stereotyped motifs include static shape or minor reaching with centroid towards rear (4); forward lengthen with centroid towards front (5); and edge centroid forward (9) (Fig. \ref{fig:figure4}b). In the run mode, stereotyped motifs include front-widen then streamline and extend (8); and retract and front-widen then extend (2). A probability matrix for transitions between the stereotyped motifs is shown in Fig. \ref{fig:figure4}d, with rows and columns corresponding to the start and end motifs, respectively. We assigned points to the closest stereotyped motif and counted transitions only once the cell moves to a different motif (i.e. diagonal entries are zero). Frequent transitions include from 3 to 7 (retract to reach to one side) and from 8 to 1 (front-widen then streamline and extend to front-widening).

\begin{figure}
    \center{\includegraphics[]
    {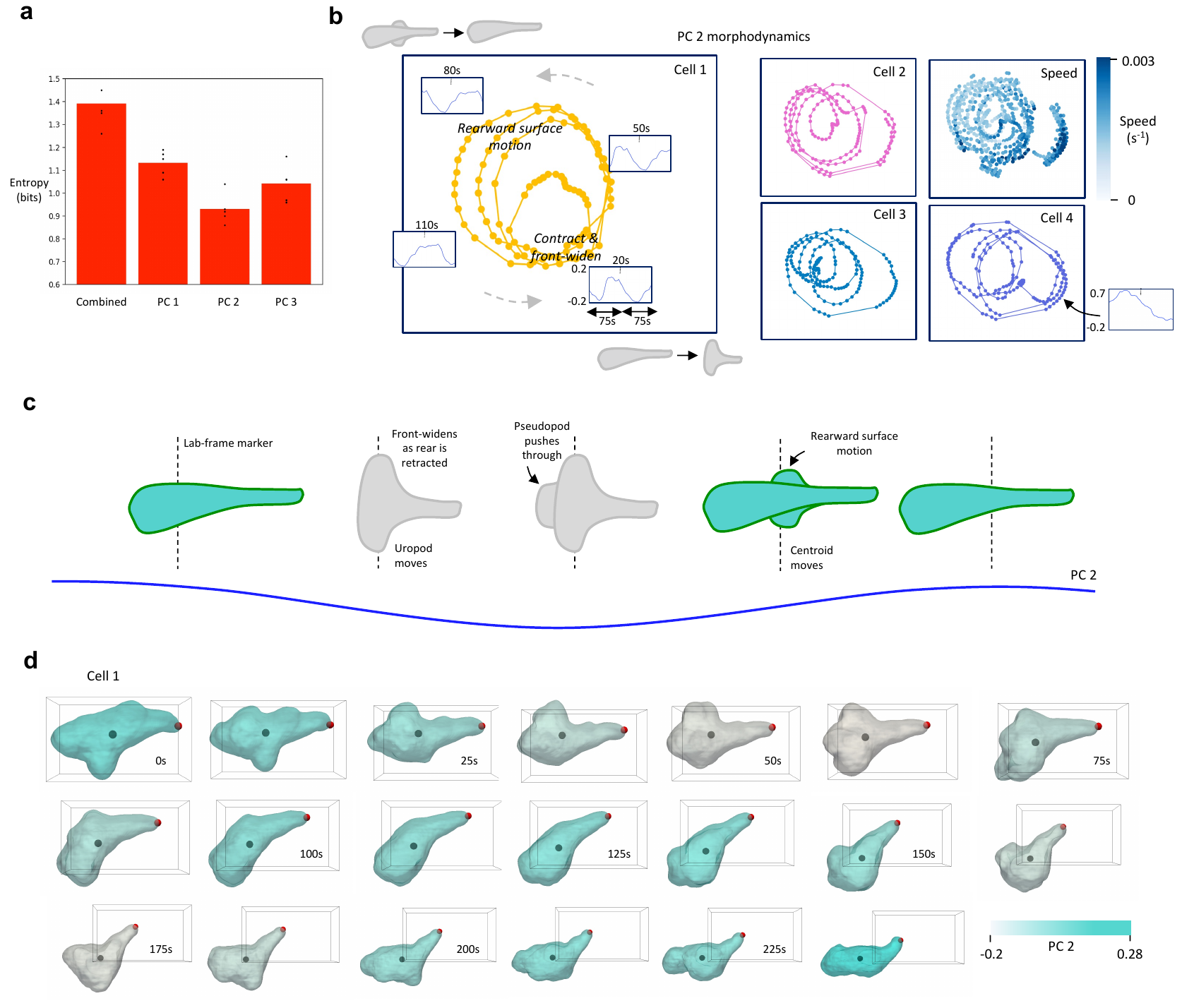}} \caption{\label{fig:figure5}\textbf{Periodic oscillations in PC 2 underlie the run mode.} \textbf{(a)} Entropy of the run mode marginal dynamics for each PC (5 repeats for each) shows a minimum for PC 2, and therefore that these dynamics are the most stereotyped, consistent with the autocorrelation results of Supplementary Fig. 5b. Markov chain entropies were calculated for transitions on grids over morphodynamic spaces for each PC, found by repeating the wavelet analysis with each PC on its own. \textbf{(b)} Dynamics in the PC 2 morphodynamic space for the run mode, where different locations represent different local PC 2 morphodynamics (and decreasing PC 2 represents contraction and front-widening). Tracking the trajectories of the longer duration datasets reveals periodic oscillations of varying amplitude. Local PC2 dynamics in 150 s windows are shown inset at key points in the morphodynamic space, showing outer rings represent higher-amplitude oscillations, with a region for particularly large PC 2 decreases, bottom right. The top left corner represents rearward surface motion and the bottom right corner represents contraction and front-widening. Regions in between represent transitions between these motifs. Maximum uropod speeds correspond to contraction and front-widening. \textbf{(c)} These results are suggestive of the following cyclic morphodynamic propulsion mechanism: the leading edge widens, likely intercalating with the ECM and contracting the uropod; the leading edge then extends forward, as the previously-widened leading edge regions undergo a rearward motion that may connect with the ECM like a paddle. This cycle repeats every $\sim$100 s. \textbf{(d)} An example showing these oscillations in Cell 1, coloured by PC 2.}
\end{figure}

We next looked in detail at the run mode, of particular interest as this is when cells use global morphodynamics for active translocation through the ECM, and because it is in all cases defined by polarised morphologies, for which our descriptor was designed. First, we repeated the wavelet analysis with the longer duration datasets of the run mode, finding more of a continuum than the global morphodynamic space, but for which stereotyped motifs can still be categorised (Supplementary Fig. 9a-b, with PC timeseries of three examples from each motif in Supplementary Fig. 10). Supplementary Fig. 9c-d shows the speeds and transition probability matrix. Aside for a turning motif, all fall into two categories: compression, and a rearward surface motion with extension forward (rearward with respect to the cell frame of reference, and relatively static in the lab frame). The precise motifs are then variants on these base behaviours, e.g. whether there is also widening. These underlying morphodynamics, omitting the distracting variations, are most characteristic of PC 2 dynamics, and this connection between PC 2 dynamics and migration is certainly consistent with the increased autocorrelation timescales and power of PC 2 relative to the stop mode (Supplementary Fig. 5b-c).

To test this theory, we calculated the entropy of each PC's morphodynamics. We did this by repeating the wavelet analysis for each PC on its own and calculating the Markov chain entropy for transitions on a grid over the resulting morphodynamic space (see Methods and Supplementary Fig. 9e). We used grids since these dynamics formed continuums rather than discrete, categorisable morphodynamic spaces. We found an entropy minimum for PC 2 (Fig. \ref{fig:figure5}a), confirming that PC 2 dynamics are the most stereotyped. Fig. \ref{fig:figure5}b shows how all four cells follow the same circular oscillations of varying radius in the space of PC 2 morphodynamics. Supplementary videos 1-4 are labelled as Cells 1-4. Outer and inner rings represent high and low amplitude oscillations, respectively, and there is a region for particularly large decreases in PC 2 (but not for large increases). These results, in conjunction with Supplementary videos 1-4, coloured by PC 2, suggest the following morphodynamic propulsion mechanism (sketched in Fig. \ref{fig:figure5}b-c): the leading edge widens, likely intercalating with the ECM to contract the uropod (PC 2 decreases); the leading edge then extends forward, as the previously-widened leading edge regions undergo a rearward flow that may connect with the ECM like a paddle, ultimately streamlining (PC 2 increases). This cycle is repeated every $\sim$100 s, and explains the oscillations in (uropod) speed observed in Fig. \ref{fig:figure3}d. Fig. \ref{fig:figure5}d shows an example section of Supplementary video 1, coloured by PC 2. These results suggest T cells utilise a highly periodic internal machinery to generate a sustained migration effort, alternating between two previously proposed propulsion mechanisms to move the uropod then leading edge forward \cite{reversat2020cellular, fowell2021spatio, abercrombie1970locomotion}. A plausible mechanistic basis for the rearward morphodynamic flow is retrograde cortical actin flow, a process that has been implicated in amoeboid migration in a number of cells, including T cells \cite{abercrombie1970locomotion, reversat2020cellular}. However, further investigations of internal actin dynamics are needed to explore this connection.

\section{Discussion}

T cells are a key part of the adaptive immune system, migrating through the extracellular matrix (ECM) to neutralise infected and cancerous cells. However, their morphodynamics have not yet been completely quantitatively mapped in 3D. Here, we used lattice light-sheet microscopy (LLSM) to acquire datasets of primary mouse cytotoxic T cells migrating through a collagen matrix with high spatiotemporal resolution. Using a novel shape descriptor that incorporates key polarisation information with a uropod label, we found that shape was low-dimensional. Run-and-stop migration emerges at long timescales ($\sim$150 s), and global morphodynamics are stereotyped, forming a discrete set rather than continuum. Stop mode morphodynamics primarily involve oscillations in centroid movement towards the uropod, with extension forwards or sideways (PC 3 dynamics), and these remain autocorrelated for long timescales (decay time, $\tau\textsubscript{ACF} \sim$250 s). The run mode (i.e. active translocation) arises from periodic oscillations in PC 2, with a period of $\sim$100 s and $\tau\textsubscript{ACF} \sim$225 s: the leading edge widens, likely using intercalation with the ECM to contract the uropod (PC 2 decreases); the leading edge then extends forward, as the previously-widened leading edge regions undergo a rearward motion that may connect with the ECM like a paddle, ultimately streamlining (PC 2 increases). These results indicate periodicity in the cellular machinery help sustain forward motion during active translocation.

Uropod tracking proved vital for differentiating key morphological and morphodynamic states. Uropod uncertainties were then required to ensure analysis was at sufficient signal-to-noise ratio (SNR), because the uropod is a diffuse region rather than a precisely-locatable point. In analogy to the role of the Hessian matrix in parameter fitting, we found this could be achieved relatively simply by quantifying uropod uncertainty through the curvature of the cell rear, then propagating this to downstream variables of interest. The inclusion of landmark-like but diffuse features will likely become more important as methods for tracking intracellular structures at high spatiotemporal resolution continue to improve, meaning spatial regions can be associated with specific internal organisation and activity \cite{mckayed2013actin}. In a small number of cases (e.g. Supplementary video 7), thin fluid-like protrusions extend out of the uropod, which cause dynamics in $D_{0}$ that are unlikely to be important for migration. To reduce these effects, in future work we will explore labelling uropods based on smoothed reconstructions (with e.g. $l_{max}=15$). We found uropod definition reduced for some cells in a long-lived stop mode (and therefore had high uncertainties for some PCs, meaning they were omitted from analyses). This may be indicative of loss of polarisation, so for these modes alternative shape descriptors may be more appropriate.

Internal retrograde actin flow has been a hallmark of cell migration models for decades, since Abercrombie first observed centripetal flow of particles on fibroblast surfaces \cite{dupre2015t, abercrombie1970locomotion}. However, Abercrombie also proposed a second propulsion mechanism, where rearward flows of surface deformation might push the cell forward like a paddle. Such morphodynamic flows (or `waves') have recently been observed in 2D migrating \textit{Dictyostelium} cells \cite{driscoll2012cell}, and in T cells embedded in microfluidic channels where they can enable migration without any adhesion \cite{reversat2020cellular}. To our knowledge, however, they have not been characterised in 3D ECM environments. Through inhibition at obstacles and activation on the opposite side, flows may also aid turning as has been described in neutrophils \cite{weiner2007actin}, and the lateral protrusions likely serve as an anchor in confined geometries \cite{tozluouglu2013matrix, mandeville1997dynamic}. Analysis of actomyosin dynamics, as well as tracking of the ECM fibres (perhaps with a contact map over the cell surfaces), would help test the connection between the rearward surface motion and internal actin dynamics, and the specific nature of how these interact with the ECM for anchoring and propulsion. The analysis would also reveal the extent to which decreasing PC2 (contraction with front-widening) is driven by contact with fibres, although the periodic PC 2 dynamics across all run mode cells suggests this may predominantly be internally regulated. 

Exciting areas for future work include the extension of the analysis to the timescale of hours, where the statistics and morphodynamics of switching between run and stop modes could be interpreted at the single-cell level, and the hierarchical organisation of the stereotyped motifs could be mapped. There are technical challenges, however: individual cells would have to be followed and migration distances would exceed the scales of current LLSM fields of view. Dataset sizes might also become problematic, given that a 20 min video corresponds to 1 TB of data (with one colour). Furthermore, non-stationary issues such as aging, differentiation and activation may come into effect \cite{metzner2015superstat}. It would also be interesting to build statistical models of T cell morphodynamics \cite{tweedy2019screening}, which may then enable the development of mechanistic models \cite{zhu2016comp}, connecting morphodynamics to both extra and intra-cellular processes.

Ultimately, we hope quantitative morphodynamic analyses of T cells navigating the complex ECM environment will aid comparison of migration across different conditions (e.g. tissues, drugs and cell mutants). In particular, the prevalence and switching between human-labelled modes of migration such as chimneying, mesenchymal, amoeboid (blebbing), finger-like, and rear-squeezing could be put on firm objective grounds \cite{zhu2016comp, yamada2019mechanisms}. These advanced morphodynamic analyses will in turn help the development of mechanistic models, with a view to enhanced understanding of, and more effective, immunotherapeutics. 

\section{Methods}

\begin{small}

\subsection{LLSM imaging and pre-processing}

\subsubsection{Lattice light-sheet microscopy (LLSM)}

LLSM experiments were either performed on a custom-built system described in \cite{geoghegan20214d}, or on a Zeiss Lattice Light Sheet 7 microscope (Zeiss, Oberkochen, Germany). OT1-Lifeact-GFP T cells \cite{galeano2020lifeact} labelled with CellTracker Deep Red dye were excited at 642 nm, OT1-mT/mG at 561 nm and plain OT1-Lifeact-GFP T cells at 488 nm. For all experiments performed on the home-built system, Point Spread Functions were measured using 200 nm Tetraspeck beads. The acquired datasets were deskewed and deconvolved using LLSpy, a Python interface for processing of LLSM data. Deconvolution was performed using a Richardson-Lucy algorithm using the PSFs generated for each excitation wavelength. Datasets acquired on the Zeiss system were deskewed using the Zeiss Zen (blue edition) software. All data were acquired at 37$^{\circ}$C and 5\% humidified CO2. The voxel size was  0.1x01x0.2 \textmu m\textsuperscript{3} for the home-built system and 0.145x0.145x0.4 \textmu m\textsuperscript{3} for the Zeiss system. The temporal resolution was 2.5 s per frame for the OT1-mT/mG datasets, 5.6 s per frame for the OT1-Lifeact-GFP-CellTracker Deep Red datasets (both imaged on the home-built system) and 4.17 s per frame for the plain OT1-Lifeact-GFP datasets imaged on the Zeiss system. We collected 29 datasets with 2,850 frames altogether and a mean and standard deviation across datasets of 98 and 78, respectively.

\subsubsection{Image segmentation}

Before further processing, membrane Tomato signal was denoised using a deep-learning approach based on Content-Aware Image Reconstruction (CARE) \cite{weigert2018content}. CellTracker Deep Red and membrane Tomato signal were bleach-corrected using FIJI \cite{schindelin9714}. Cell surfaces were segmented using Imaris 8.4.1 (Bitplane, Zurich, Switzerland). To minimise the occurrence of holes in the surfaces, depending on the signal to noise ratio smoothing factors between 0.35 \textmu m and 0.8 \textmu m were applied. Cell surface triangulations were exported using custom Matlab code and again analysed for surface holes. If required, surface holes were eliminated by custom Matlab code based on closing operations. 

\subsubsection{Sample preparation}

Primary murine OT1-Lifeact-GFP and OT1-mT/mG cytotoxic T cells were isolated and cultured as previously described \cite{galeano2020lifeact}. All imaging was done with T cells cultured over 6 or 7 days. For imaging on the home-built system, OT1-Lifeact-GFP T cells were labelled with 100 nM CellTracker Deep Red dye (ThermoFisher Scientific, Waltham, USA).
Keeping all components on ice, collagen matrix solution was prepared by adding 10 \textmu l of 10x PBS, 1.15 \textmu l 1N NaOH and 39 \textmu l T cell medium (TCM), consisting of phenol-free RPMI 1640, 10\% foetal calf serum, 1 mM sodium pyruvate, 10 mM HEPES, 100 U/ml penicillin, 100 \textmu g/ml streptomycin, 2 mM L-glutamine and  50 \textmu M $\beta$2-mercaptoethanol (all from Gibco, ThermoFisher Scientific, Waltham, USA), to 50 \textmu l liquid-phase rat-tail collagen I ($\sim$3 mg/ml; Corning, New York, USA). Coverslip and imaging dish glass surfaces were treated with 2\% (3-aminopropyl) triethoxysilane in ethanol and 6\% glutaraldehyde to facilitate firm attachment of collagen gels. For imaging on the home-built LLSM, 6 \textmu l of collagen mix were placed onto surface-treated round 5 mm coverslips (Warner Instruments, Hamden, USA) and polymerised at 37$^{\circ}$C for 15 min. After polymerisation, 10\textsuperscript{5} T cells in phenol-free TCM were seeded on top of the gel and allowed to infiltrate over 3 h before imaging. For imaging on the Zeiss LLS system, 10\textsuperscript{5} T cells were added to TCM during collagen matrix mix preparation. 70 \textmu l of collagen mix were added to well of 35 mm imaging dishes (Mattek, Ashland, USA) and polymerised at 37$^{\circ}$C for 30 min. After polymerisation, 1 ml of pre-warmed phenol-free TCM was added to the dish and cells were allowed to recover for 1 h before imaging.

\subsection{Quantifying 3D cell morphology}

Cell morphologies were quantified using SPHARM. First, the cell surface, described with 3 Cartesian coordinates, $\{x, y, z\}$, is mapped to the unit sphere, described with polar coordinates $\{\theta, \phi\}$, such that the three Cartesian coordinates are functions of the polar coordinates: $\{x(\theta, \phi), y(\theta, \phi), z(\theta, \phi)\}$.  $\{x(\theta, \phi), y(\theta, \phi), z(\theta, \phi)\}$ are then be decomposed in terms of the spherical harmonics, $Y_{l}^{m}(\theta, \phi)$, and only $m\geq0$ functions are required \cite{styner2006framework}. For $x$ for example, 
\begin{equation}
    x(\theta, \phi) = \sum_{l=0}^{\infty}\sum_{m=0}^{l}c_{l, x}^{m}Y_{l}^{m}(\theta, \phi),
    \label{eq:decomposition}
\end{equation}
and the (in general complex) coefficients, $c_{l, i}^{m}$ with $i\in \{x, y, z\}$, represent the morphology. We used the SPHARM-PDM software package \cite{styner2006framework} to find the coefficients for the T cells with $l_{max}=15$ and cell meshes converted to voxel grids with a spatial resolution of 0.5 $\mu m$ for computational speed. The additional variable for capturing polarisation information was $D_{0}$. This was the distance between the uropod and centroid multiplied by $\frac{3}{2}$, with the numerator reflecting the fact that the harmonics are summed over 3 spatial coordinates and the denominator accounting for the fact that the coefficients have a spatial extent double their magnitude. The uropod was manually selected (aiming for its center) in alternating frames and linearly interpolated. PCA is a dimensionality reduction method that finds a set of uncorrelated linear features (the principal components, PCs), which are the eigenvectors of the data covariance matrix (which for $D_{l}$ has dimensions 16 $\times$ 16) \cite{wold1987principal}. Supplementary Fig. 2c shows the vector composition of each PC. As explored through the main text, PC 1 is largely associated with transitions between run and stop mode morphologies, PC 2 is largely associated with morphological transitions in the run mode, and PC 3 is largely associated with morphological transitions in the stop mode. For implementing PCA, we used the Scikit-learn Python package.

\subsection{Uncertainty quantification}

The uncertainty in the uropod label depends on the curvature of the cell rear, which we quantified using the mean curvature averaged across the 15 closest mesh vertices to the labelled point (with a sub-sampled mesh for computational speed). We then defined the positional uncertainty as the cord length associated with a 20$^{\circ}$ rotational perturbation. To convert this to PC uncertainties, we found the set of possible $D_{0}$ values using mesh vertices within this uncertainty (i.e. within one cord length of the uropod label), calculated the standard deviation, and converted to PC uncertainties by multiplying by the cosine of the angle between the $D_{0}$ and PC vectors in $\{D_{l}\}$ space. This process was repeated for every 10 frames in each dataset to get a single characteristic uncertainty for each PC (the mean) for each dataset. $\tau_{sig}$ was calculated as the mean time take for the uropod to move twice the cord length. Some cells have uropods that are near-stationary, and therefore have a $\tau_{sig}$ comparable with the full dataset duration. To account for such cases, we used a maximum $\tau_{sig}$ of 100 s, in order that these could be plotted for comparison with dynamic cells, but we excluded them from quantitative analysis.

\subsection{Finding a motion variable for small timescales}

We calculated uropod and centroid velocities by finding the displacements between consecutive positions (smoothed with running means over $\tau_{sig}$ for both, for consistency) and dividing by the time step and cube root of cell volume (for invariance to cell scale). The ellipsoid major axis was calculated as the eigenvector with the largest eigenvalue of $A^{T}A$ where A is a matrix of the $l=1$ spherical harmonic coefficients \cite{brechbuhler1995parametrization}: $\frac{\sqrt{3}}{3\sqrt{2\pi}}(\mathbf{c}_{1}^{-1}-\mathbf{c}_{1}^{1}, i(\mathbf{c}_{1}^{-1}-\mathbf{c}_{1}^{1}), \sqrt{2}\mathbf{c}_{1}^{0})$. For comparing the uropod-centroid (UC) and ellipsoid axes, we used running means for uropod and centroid with a time window of 100 s for long-timescale behaviour. We compared cells where the uropod speed was above 0.0025 s$^{-1}$, i.e. moving more than a quarter cell length in 100 s, and the distance between the uropod and centroid velocity vectors was within half the uropod speed, i.e. they were aligned.

\subsection{Timeseries autocorrelation functions and power spectra}

Autocorrelation functions and power spectra were computed for longer duration datasets for each of the run and stop mode. We removed timeseries with low SNR: PC timeseries where the ratio of the signal standard deviation to the PC uncertainty was below 2.5 and speed timeseries where $t_{sig}$ was of a similar scale to the full dataset duration. There was one removal for each of PC 1, 3 and speed, across different cells. We calculated the autocorrelation on de-trended timeseries, in order to only capture statistically significant correlations, removing trends with frequencies lower than 0.0025 Hz (corresponding to a period of approximately half the total dataset duration) with a Butterworth highpass filter \cite{butterworth1930theory}. We then found a decay time, $\tau\textsubscript{ACF}$, by fitting an exponential decay model, $y=e^{-\frac{x}{\tau\textsubscript{ACF}}}$,  to the peaks of the ACF (rather than the full ACF, which is more appropriate for non-oscillatory patterns).
 
\subsection{Continuous wavelet transform}

The continuous wavelet transform was used to find local morphodynamics (or `behaviours') from the PC timeseries. A wavelet that decays to zero either side of a central peak is convolved with the timeseries, which produces a new timeseries where each element now represents local morphodynamics. Repeating this process with dilated versions of the wavelet and stacking the resulting set of timeseries yields a spectrogram with multiscale dynamic information, where high-frequency components are analysed close in time, but lower frequency information bleeds in from afar. This spectrogram is then mapped to an interpretable 2D space using t-SNE \cite{van2008visualizing}, and a PDF can be computed with kernel density estimation \cite{davis2011remarks}. t-SNE (t-distributed stochastic neighbour embedding) is a non-linear dimensionality reduction method that uses machine learning. Two similarity metrics between datapoints are defined for each of the two representations, the initial (high-dimensional) representation and the target (lower-dimensional) representation. The difference between the distributions of these similarities across all data pairs is minimised. For implementing t-SNE, we used the Scikit-learn Python package with default parameters: perplexity (analogous to the number of neighbours in other algorithms) of 30, learning rate of 200, and 1000 iterations. 

We identified stereotyped motifs (PDF peaks) using adaptive binarisation, a method that thresholds pixels in an image with a threshold value that depends on the local statistics: the mean over a surrounding square of pixels with an added bias (we used square dimensions of 7 and a bias of 20, found with a grid search). We used adaptive rather than pure binarisation so that regions with high-density peaks and high PDF between then (`superhighways' representing common transitions) could be separated, while lower peaks in absolute terms could also be captured. We used two simple wavelets, the `mexican hat' wavelet and Gaussian 1\textsuperscript{st} derivative wavelet, with the combination of the two required to capture symmetric and antisymmetric features. For organisms where the morphodynamics of interest are organised in repeating bouts, e.g. high-frequency wing-beating of fruit flies, complex wavelets that enable the removal of phase information can be useful. However, over the timescales analysed here, T cell morphodynamics are slower-changing, and phase information is important. We used six equally-spaced frequencies for each wavelet from double the Nyquist limit up to the (wavelet-specific) frequency with width of influence corresponding to 150 s, the approximate timescale of organisation found from the autocorrelation analysis. The width of influence was found by convolving each wavelet with a square pulse to find where edge effects begin. When repeating this method for only the four run mode datasets, we used for the adaptive binarisation parameters square dimensions of 15 and a bias of 50, again found with a grid search.

\subsection{Comparing marginal morphodynamics of the run mode}

The marginal morphodynamics form continuums, and so transition matrices over stereotyped PDF peaks cannot be defined. Instead, we defined transition matrices over points on a grid. We then quantified the entropy for the transition dynamics of each PC (and compared with that of their combined dynamics). The entropy is $-\sum{\pi_{i}p_{ij}log_{2}p_{ij}}$, where $\pi_{i}$ is the equilibrium distribution and $p_{ij}$ is the probability that the next motif to be visited after $i$ will be $j$. For plotting the PC 2 dynamics of the four cells, we perturbed the wavelets slightly to further improve the interpretability. This was done by searching locally across options for the maximum wavelet width (keeping 150 s as an upper bound) and finding combinations with reduced entropy. Reduced entropy was associated with reducing the Gaussian wavelet maximum width to 100 s, but with the same Mexican hat wavelets as before.

\end{small}

\vspace{5mm}

\noindent \textbf{Author Contributions} \\
DK, JM and MB acquired and segmented the cell surface segmentation data; HC and RE performed the downstream morphological and morphodynamic analysis. \\

\noindent \textbf{Data accessibility} \\
The cell surface segmentation data that support the findings of this study have been deposited on Dryad\\
(DOI: \url{https://doi.org/10.5061/dryad.tdz08kq1r}). The code used is available at \url{https://github.com/hcbiophys/tcells_paper_code}. \\

\noindent \textbf{Funding Statement} \\
This work was funded by the Biotechnology and Biological Sciences Research Council (grant number BB/M011178/1) to RE and the Australian Research Council (Discovery Project grant DP180102458) to MB.\\

\noindent \textbf{Acknowledgements} \\
The authors thank ND Geoghegan, KL Rogers, N Tubau and LW Whitehead (WEHI, Melbourne, Australia) for technical assistance with LLS microscopy and image denoising. MB acknowledges Bitplane AG for an Imaris Developer licence, and HC and RE thank Suhail Islam for invaluable computational suppport.  \\

\noindent \textbf{Competing Interests} \\
The authors declare no competing interests. \\

\printbibliography

\end{document}


\pagenumbering{gobble} 
\maketitle
\begin{center}
Henry Cavanagh\textsuperscript{1}, Daryan Kempe\textsuperscript{2}, Jessica K. Mazalo\textsuperscript{2}, Maté Biro\textsuperscript{2}, Robert G. Endres\textsuperscript{1*} \\
*Corresponding author: r.endres@imperial.ac.uk
\end{center}

\noindent \textsuperscript{1}Imperial College London, Centre for Integrative Systems Biology and Bioinformatics, London, UK, SW7 2BU; \textsuperscript{2}EMBL Australia, Single Molecule Science node, School of Medical Sciences, The University of New South Wales, Sydney, Australia.

\tableofcontents
\newpage

\section{Supplementary Figures}
\pagenumbering{arabic}  

\subsection{Supplementary Figure 1: 3D T Cell Migration}

\begin{figure}[H]
    \center{\includegraphics[]
    {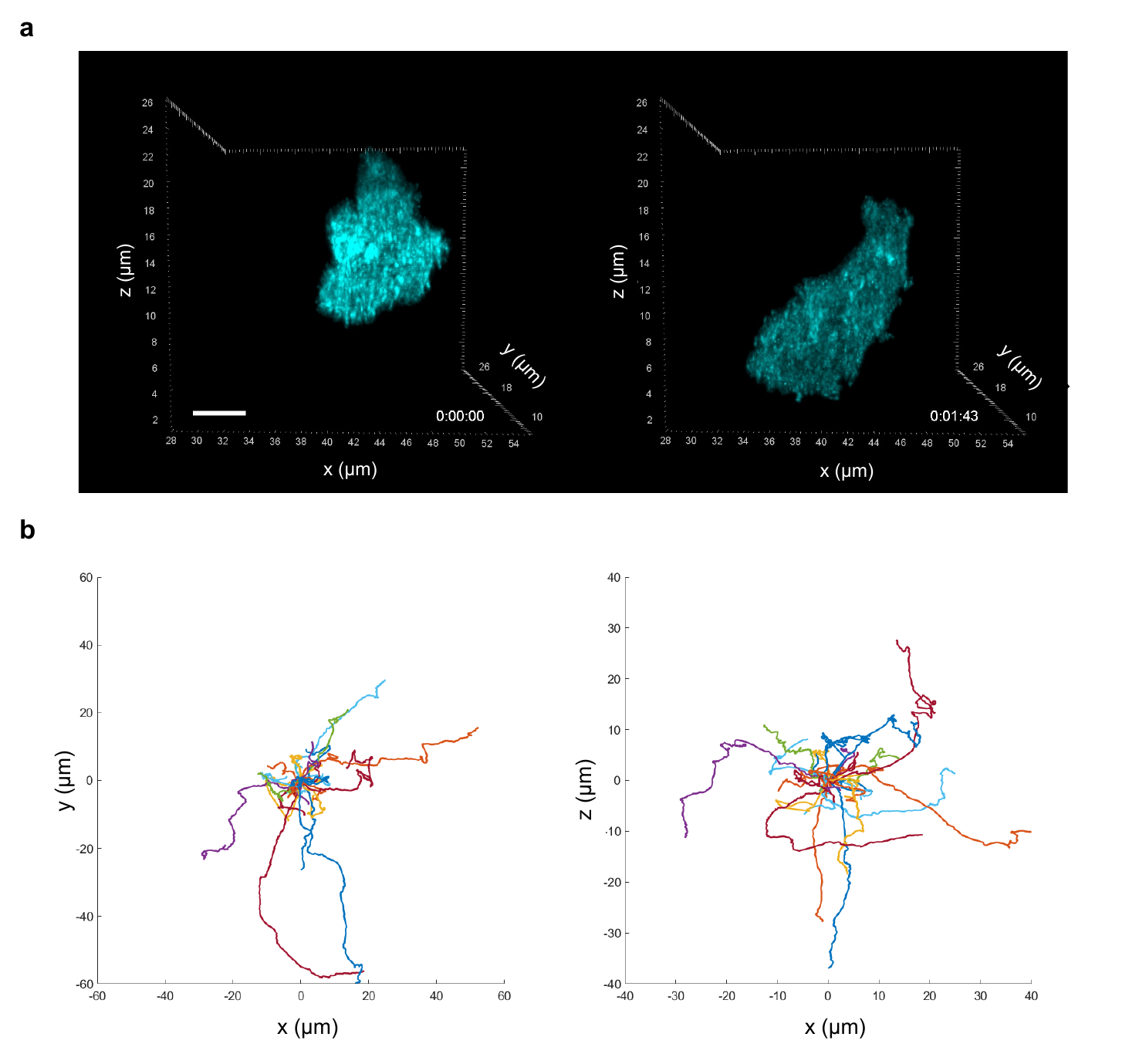}}        
    \caption{\label{fig:3d} \textbf{3D T Cell Migration.} \textbf{(a)} Representative snapshot of a T cell migrating in a 3D collagen gel. Scale bar: 5 \textmu m. \textbf{(b)} Migration tracks of T cells embedded in a 3D collagen gel. Left: xy view. Right: xz view. Starting point of tracks was translated to origin of coordinate system for visualisation purposes.}
\end{figure}

\subsection{Supplementary Figure 2: Full meshes and principal components of the sampled frames}

\begin{figure}[H]
    \center{\includegraphics[]
    {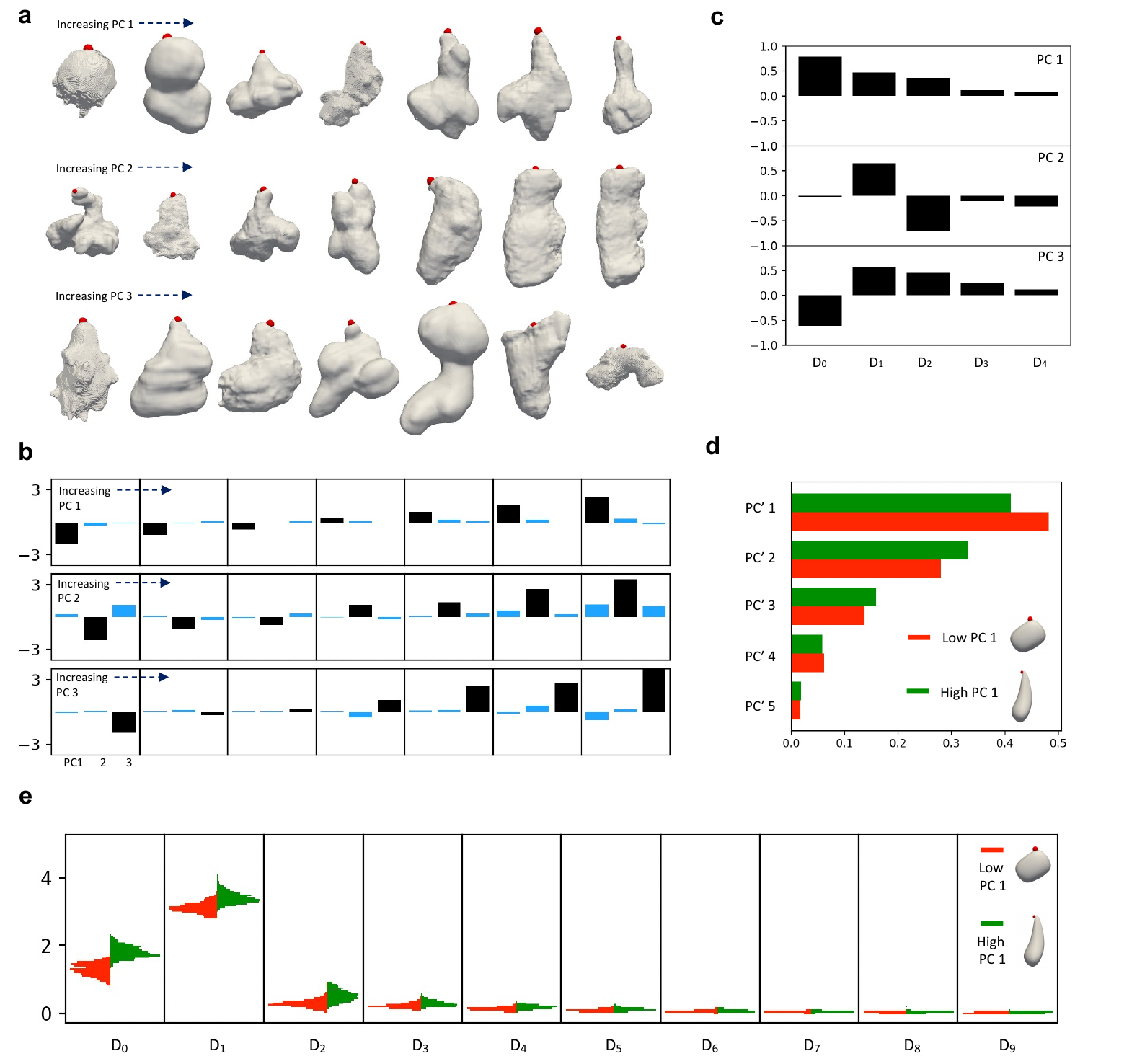}}        
    \caption{\label{fig:SI_pca} \textbf{Full meshes and principal components of the sampled frames.} \textbf{(a)} Shape changes associated with each PC are shown, found by splitting the PCA space into 7 equal-length bins along each axis and plotting the T cell within each bin with the lowest value for the other PCs ($l_{max}=3$ reconstructions in the main text, full cells here). An increasing PC 1 represents elongation and front-widening, a decreasing PC 2 represents contraction with front-widening, and an increasing PC3 represents elongation (forward or sideways), with the centroid moving towards the uropod. \textbf{(b)} Normalised PC values of the displayed cells, where black colouring indicates which PC is being sampled. \textbf{(c)} Vector composition of each PC, in terms of the spherical harmonic descriptors, $D_{l}$. \textbf{(d)} Dimensionality across the main mode of variation (PC 1) is relatively constant. For data both below the mean along PC 1 (red) and above it (green), the explained variance ratios by a new set of PCs, PC$'$, are plotted, and these decay at similar rates. \textbf{(e)} Difference in the spherical harmonic spectra (expressed through the descriptors, $D_{l}$) between the low and high PC 1 (below and above the PC 1 mean, respectively) populations.}
\end{figure}

\subsection{Supplementary Figure 3: Uropod uncertainty quantification and propagation to downstream variables of interest}

\begin{figure}[H]
    \center{\includegraphics[]
    {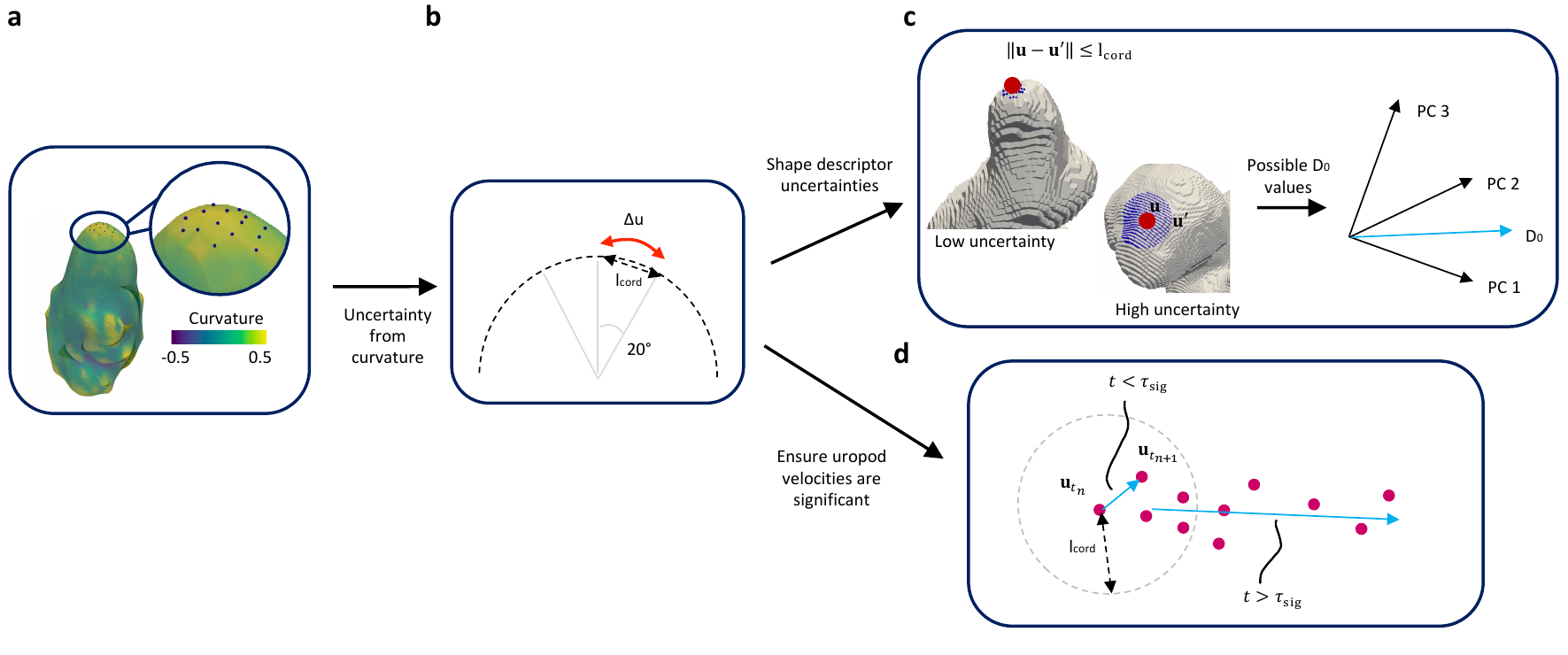}}        
    \caption{\label{fig:SI_uncertainty} \textbf{Uropod uncertainty quantification and propagation to downstream variables of interest.} \textbf{(a)} Mean curvature across the closest 15 mesh vertices to the uropod label was calculated (with a sub-sampled mesh for computational speed). \textbf{(b)} Uropod uncertainty is defined as the cord length associated with a 20$^{\circ}$ rotational perturbation. \textbf{(c)} PC uncertainties for each video were found by recalculating $D_{0}$ with all mesh vertices within this uropod uncertainty, calculating the standard deviation, and multiplying this by the cosine of the angle between the $D_{0}$ and PC vectors in $\{D_{l}\}$ space for every 10 frames in each video, then calculating the mean. \textbf{(d)} Uropods were also tracked for connecting morphodynamics to motion. To ensure these are at sufficient signal-to-noise ratio, we found for each video the mean time taken for the uropod to move a significant distance (defined as twice the cord length), $\tau_{sig}$, and then computed velocities using uropod running means with a time window of $\tau_{sig}$. We used a maximum $\tau_{sig}$ of 100s to ensure near-stationary cells were included in visualisations, but these insignificant velocities were omitted from quantitative analysis.}
\end{figure}

\subsection{Supplementary Figure 4: Finding a motion variable for small timescales}

\begin{figure}[H]
    \center{\includegraphics[]
    {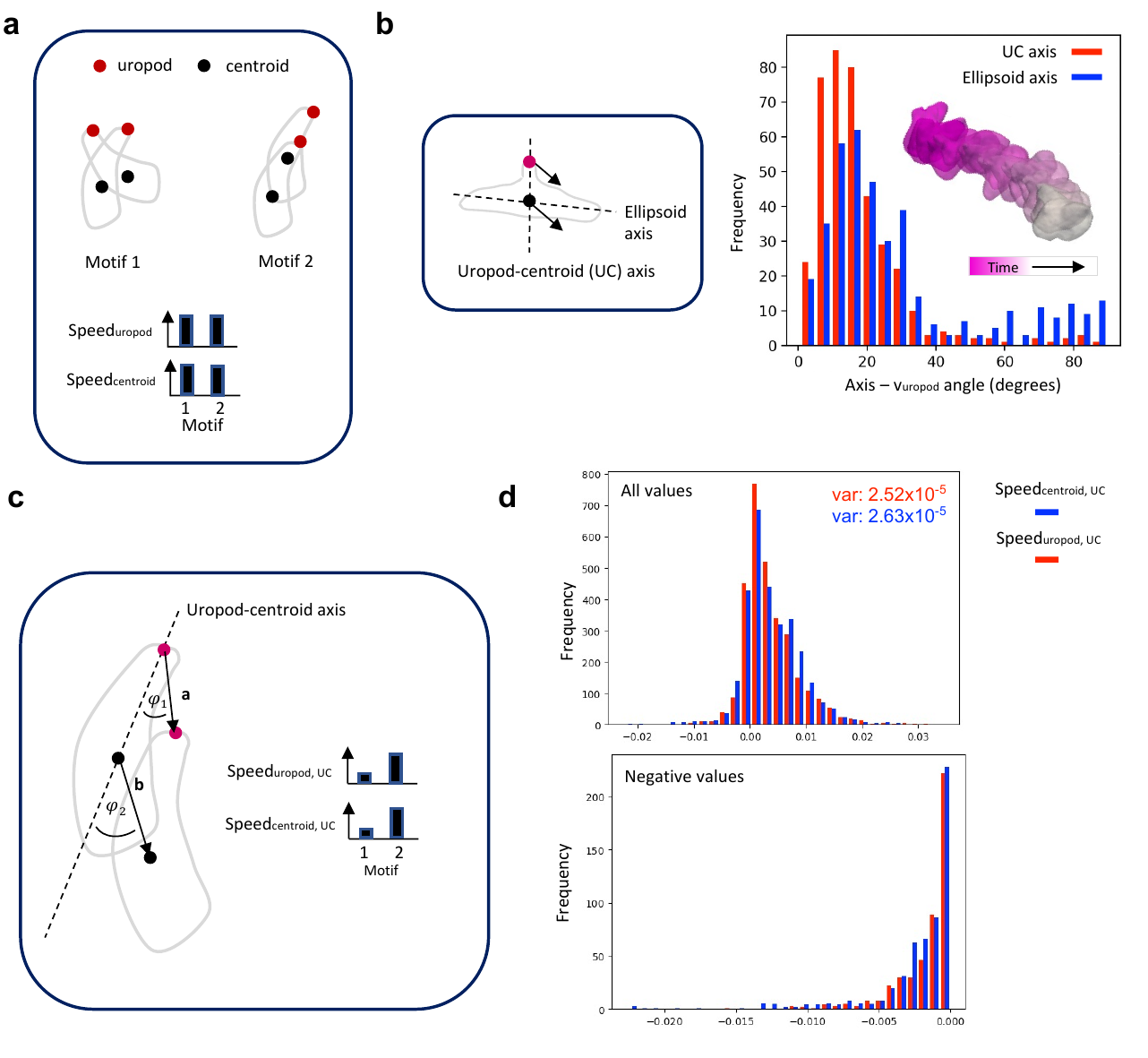}}        
    \caption{\label{fig:SI_speed} \textbf{Finding a motion variable for small timescales.} \textbf{(a)} We tracked two variables that can be used to link morphodynamics with motion: the uropod and the centroid. Their velocities (dividing by the cube root of volume for scale invariance) are 3D and not rotationally invariant, and a simple description in terms of the speed of either does not adequately separate distinct motifs, like turning (motif 1) and moving forward (motif 2). An internal reference frame is needed. \textbf{(b)} There are two options: the centroid-uropod (UC) axis and ellipsoid axis. A histogram shows that at times when the whole cell moves in unison, this happens more along the UC axis than the ellipsoid axis, with an example cell where the two differ shown. Running means for uropods and centroids with a time window of 100s were used for long-timescale behavior. Motion in unison was taken to be when uropod speed was above 0.0025 s$^{-1}$, i.e. moving more than a quarter cell length in 100s, and the distance between the uropod and centroid velocity vectors was within half the uropod speed, i.e. they were aligned. \textbf{(c)} Descriptors in terms of the speed of the uropod and centroid along the UC axis (speed\textsubscript{uropod,UC} and speed\textsubscript{centroid,UC}) can then differentiate motifs 1 and 2. \textbf{(d)} These describe largely irreversible motion (now with running means using the time windows from the uncertainty analysis, to include shorter-timescale behavior). speed\textsubscript{uropod, UC} has lower variance and fewer reversals, and Supplementary video 9 and Fig. 1f show examples where speed\textsubscript{centroid, UC} is much more oscillatory. We therefore selected speed\textsubscript{uropod, UC} as the cell motion variable, and referred to it simply as speed.}
\end{figure}

\subsection{Supplementary Figure 5: Link between run-and-stop modes and shape}

\begin{figure}[H]
    \center{\includegraphics[]
    {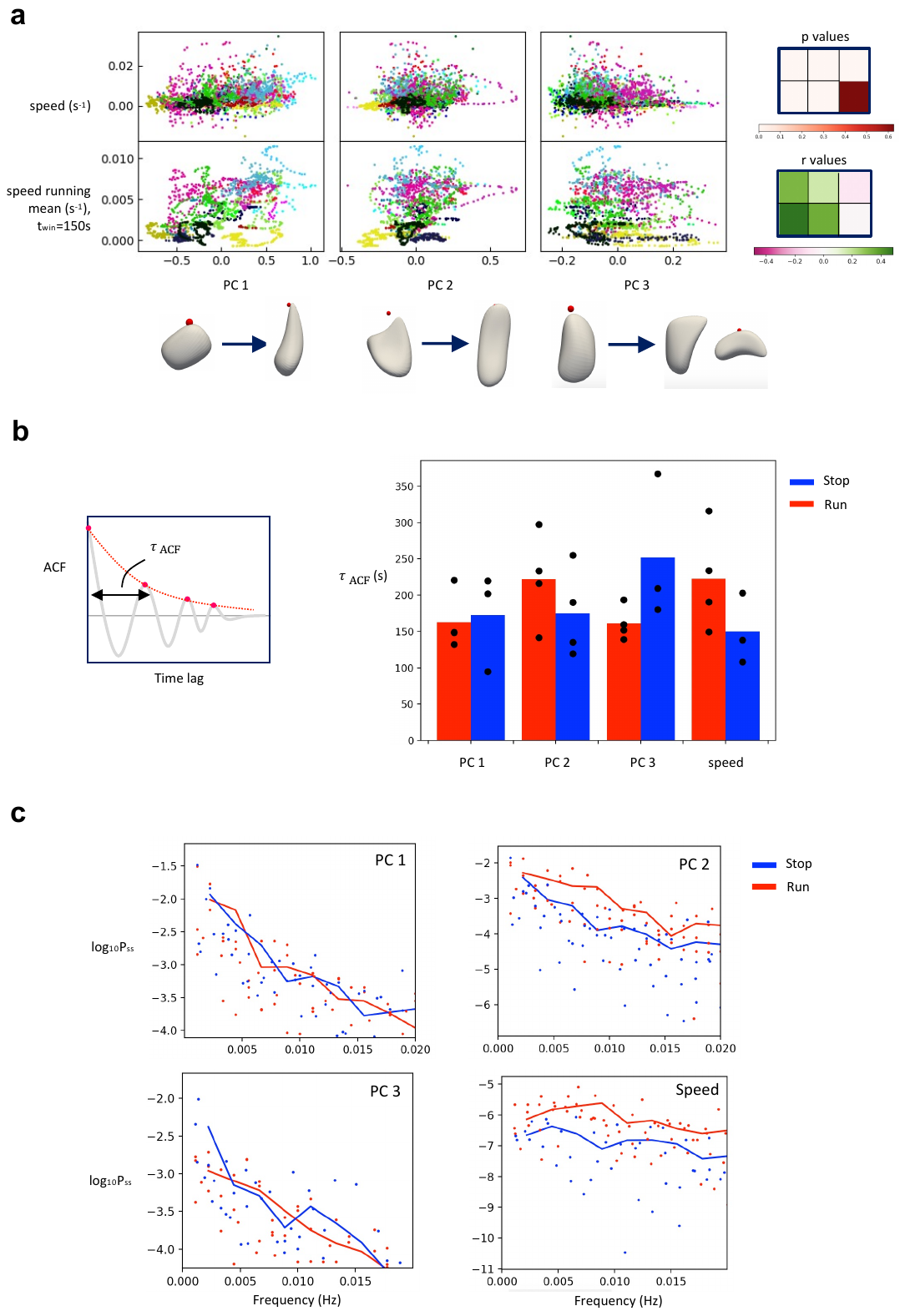}}        
    \caption{\label{fig:SI_correlations} \textbf{Link between run-and-stop modes and shape.} \textbf{(a)} The link between cell shape and speed is shown through correlations between the shape PCs, raw speed, and the running mean of speed with a window size of 150s. PCs 1 and 2 have a stronger correlation with long rather than short-timescale speed, indicating that shape is specialised more for migration mode than instantaneous speed, with cells in the run mode longer and thinner than those in the stop mode. $p$-values and $r$-values (Pearson correlation coefficients) are shown. \textbf{(b)} The autocorrelations (ACFs) were calculated for four long videos from each of the stop and run modes. Decay timescales, $\tau\_{ACF}$, were found using exponential decay models fitted to the peaks of the oscillating ACFs. For cells in the stop mode, PC 3 is the most strongly autocorrelated, followed by PC 2. For cells in the run mode, the main differences are a large drop in the autocorrelation of PC 3, meaning PC 2 becomes the most autocorrelated shape variable, and an increase in the autocorrelation of speed. \textbf{(c)} The power spectra of the PC and speed time series. The run mode is also associated with larger oscillations in PC 2 and speed. Only powers above the mean variance of associated with PC uncertainty from the uropod labelling are shown.}
\end{figure}

\subsection{Supplementary Figure 6: Continuous wavelet transform used to map stereotyped motifs}

\begin{figure}[H]
    \center{\includegraphics[]
    {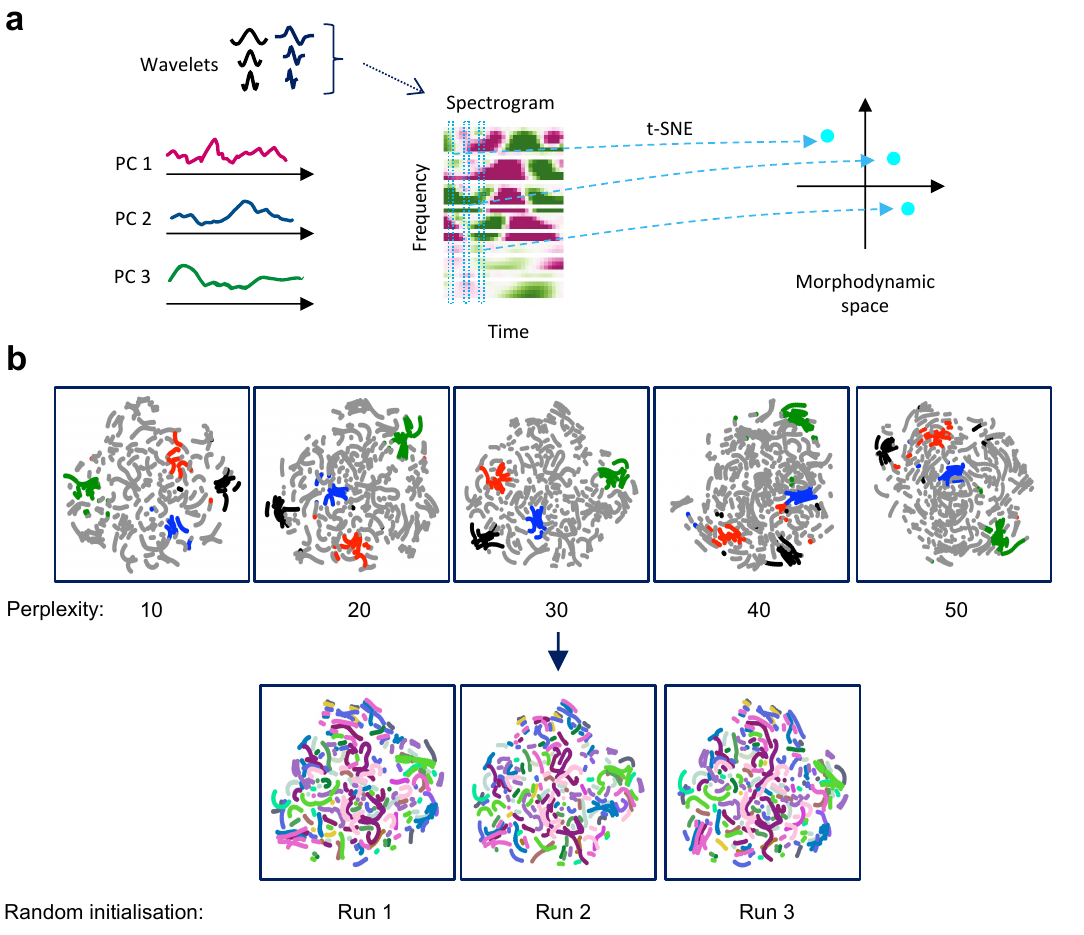}}        
    \caption{\label{fig:SI_wavelets_method} \textbf{Continuous wavelet transform used to map stereotyped motifs.} \textbf{(a)} The continuous wavelet transform was used to acquire a spectrogram capturing local multi-scale dynamic information from the PC time series, with two wavelet types to ensure both symmetric and antisymmetric features are captured (`mexican hat' and Gaussian 1\textsuperscript{st} derivative). We used 6 frequencies per wavelet, from double the Nyquist limit up to the frequencies associated with widths of influence of approximately 150s, as found from the autocorrelation analysis to be the timescale of morphodynamic organisation. The spectrogram, which represents morphodynamics at each time point, was embedded in an interpretable 2D morphodynamic space using t-SNE. \textbf{(b)} Robustness of the morphodynamic space found with t-SNE. The perplexity parameter is analogous to the number of neighbours in alternative dimensionality reduction methods, and the value used was 30 (the default for the Scikit-learn Python package). Colouring four of the motifs (top) shows that the embeddings are similar across the perplexity range suggested in the original t-SNE paper \cite{van2008visualizing}. Re-running the algorithm with a perplexity of 30 coloured by cell (bottom) also shows the embeddings are robust across random initialisations. }
\end{figure}

\subsection{Supplementary Figure 7: Global morphodynamic space with further examples.}

\begin{figure}[H]
    \center{\includegraphics[width = 0.9\linewidth]
    {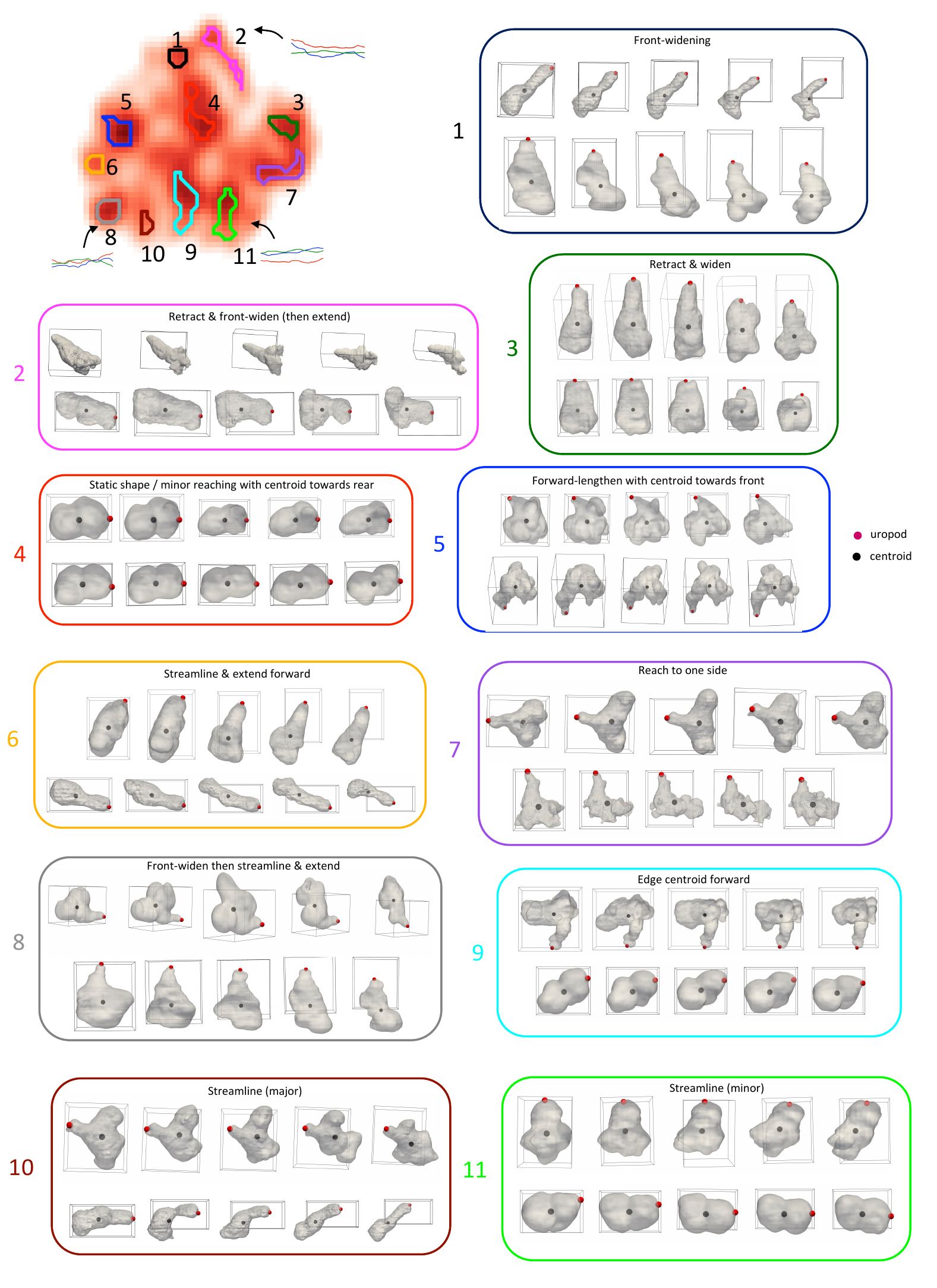}}        
    \caption{\label{fig:SI_2_motifs} \textbf{Two examples for each stereotyped motif from Fig. 4a.} } 
\end{figure}

\subsection{Supplementary Figure 8: PC dynamics of the stereotyped motifs}

\begin{figure}[H]
    \center{\includegraphics[width = \linewidth]
    {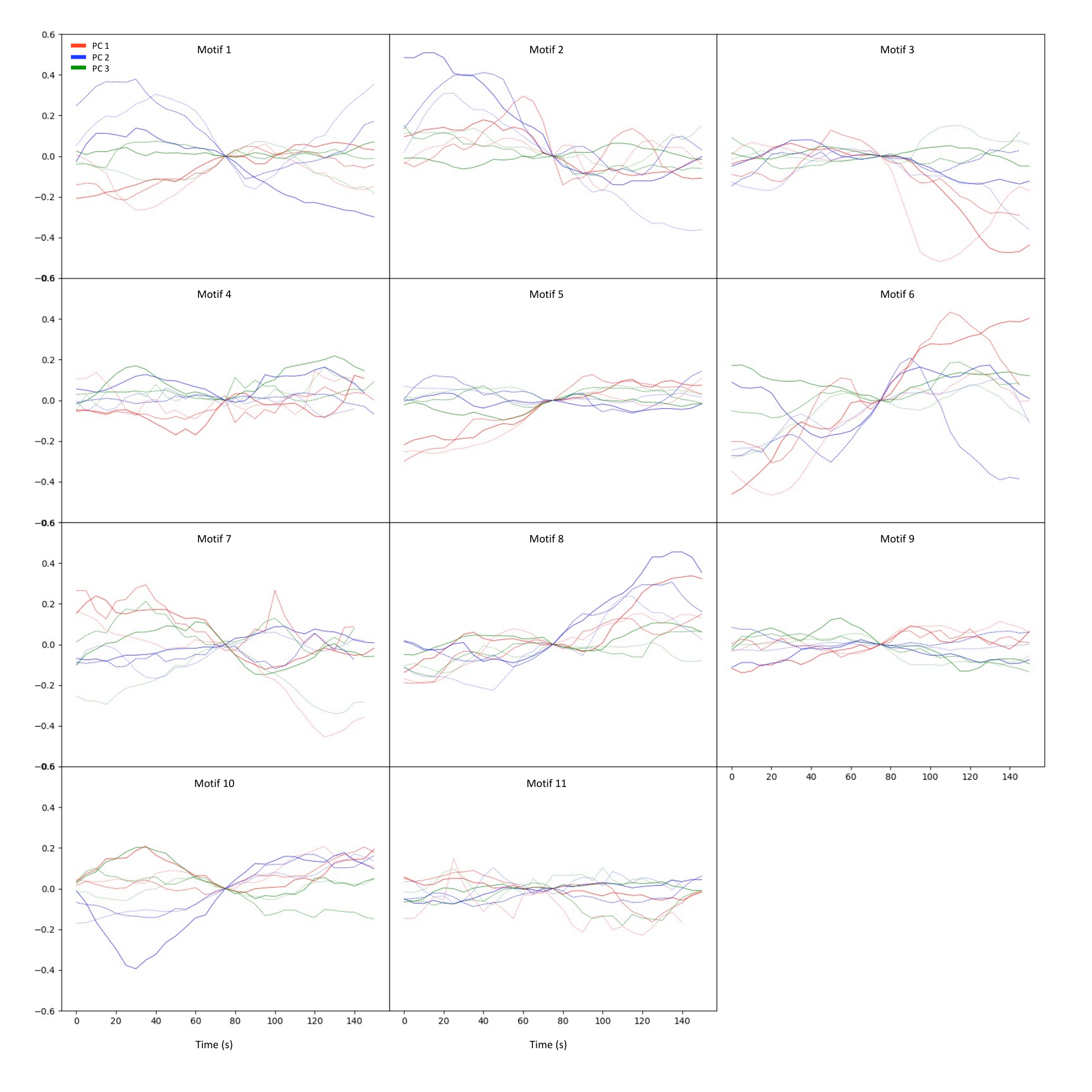}}    
    \caption{\label{fig:SI_local_PCs_combined} \textbf{PC dynamics of the stereotyped motifs.} Three principal component (PC) time series for each of the stereotyped motifs are shown, with a 150s time window, and aligned in the $y$ direction so the middle times coincide. Colours indicate the PC and different PC series have different line styles.}
\end{figure}

\subsection{Supplementary Figure 9: Run mode morphodynamics}

\begin{figure}[H]
    \center{\includegraphics[]
    {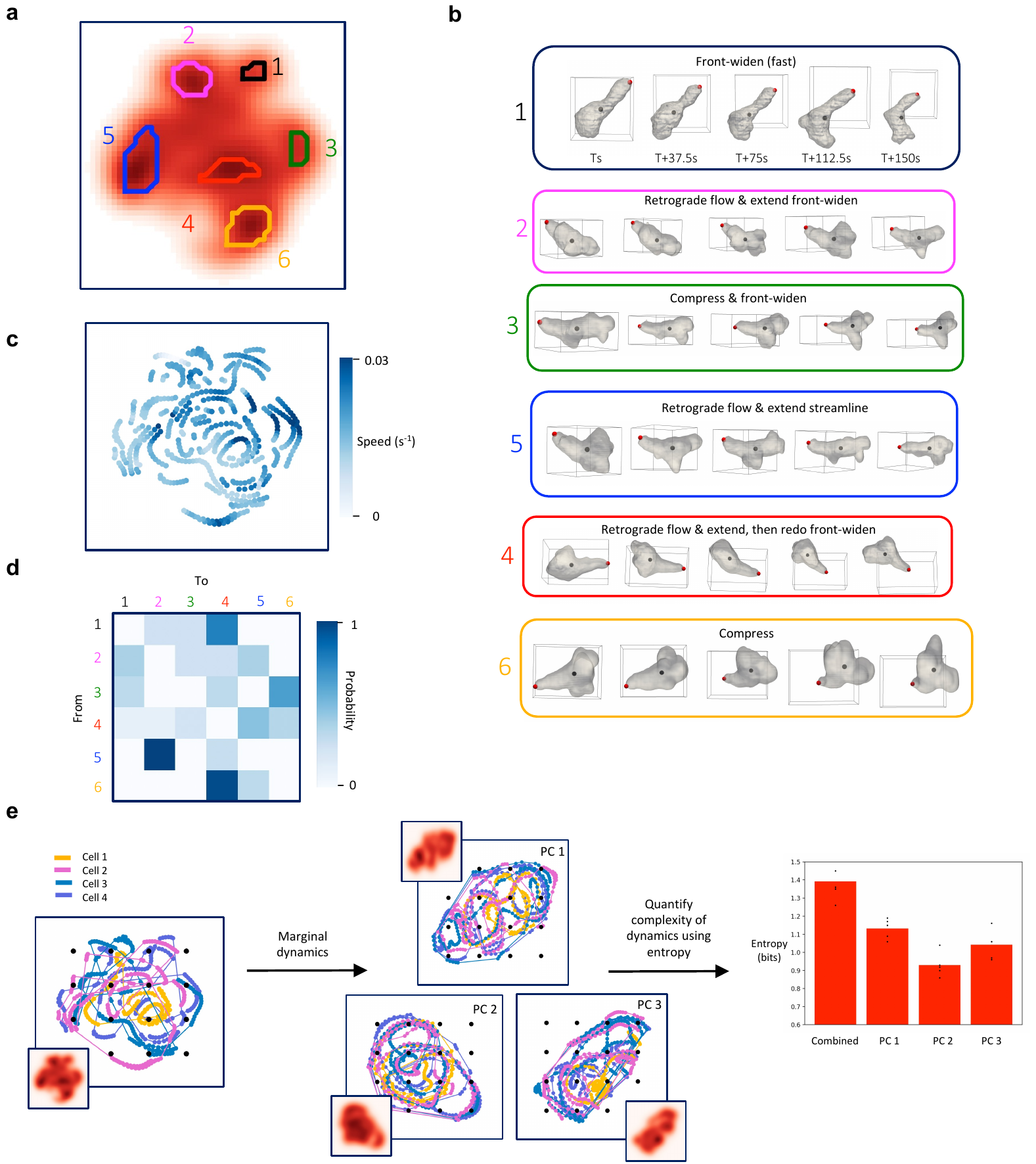}}        
    \caption{\label{fig:SI_wavelets_run} \textbf{Run mode morphodynamics.} \textbf{(a)} A morphodynamic space now formed exclusively from the four long videos of cells in the run mode shows the morphodynamic composition with higher resolution. \textbf{(b)} Examples of the stereotyped motifs, each over a 150s period. \textbf{(c)} Colouring by speed shows that motif 3 (compress and front-widen) is associated with the highest speeds. \textbf{(d)} The transition probability matrix (sequential stereotyped peaks). \textbf{(e)} Marginal dynamics of each PC. These form continuums, and so transition matrices were defined over points on a grid. We then quantified the entropy for the dynamics of each PC (and compared with that of their combined dynamics) and found an entropy minimum in PC 2 (consistent with the autocorrelation results). The entropy is $-\sum{\pi_{i}p_{ij}log_{2}p_{ij}}$, where $\pi_{i}$ is the equilibrium distribution and $p_{ij}$ is the probability that the next motif to be visited after $i$ will be $j$.}
\end{figure}

\subsection{Supplementary Figure 10: PC dynamics of the stereotyped motifs in the run mode}

\begin{figure}[H]
    \center{\includegraphics[]
    {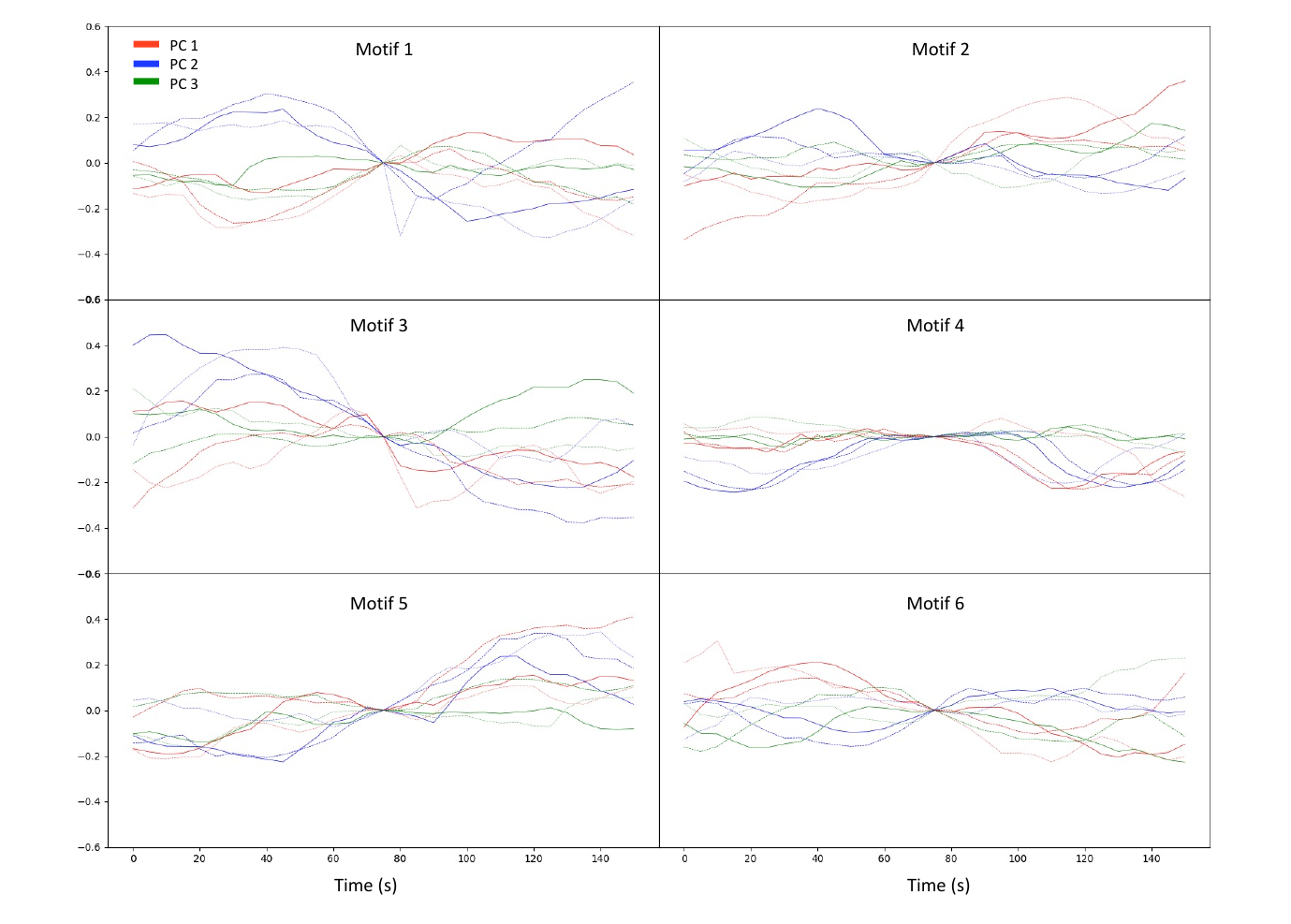}}    
    \caption{\label{fig:SI_local_PCs_split} \textbf{PC dynamics of the stereotyped motifs in the run mode.} Three principal component (PC) time series for each of the stereotyped motifs are shown, with a 150s time window, and aligned in the $y$ direction so the middle times coincide. Colours indicate the PC and different PC series have different line styles.}
\end{figure}

\printbibliography